\documentclass[prd,aps,noshowpacs,nofootinbib, 10 pt]{revtex4}
\usepackage{amsmath,graphicx,color,epsfig}

\begin{document}

\title{Scalar mesons in weak semileptonic  decays of $B_{(s)}$ }

\author{Yu-Ming Wang$^{1}$} \author{M. Jamil\ Aslam$^{1,2}$} \author{Cai-Dian L\"{u}$^{1}$}

\vspace*{1.0cm}

\affiliation{$^{1}$Institute of High Energy Physic, P. O.\ Box
918(4), Beijing,
100049, P.\ R. China. \\
$^{2}$COMSATS Institute of Information Technology Islamabad,
Pakistan}

\vspace*{1.0cm}

\date{\today}

\begin{abstract}
The transition form factors of  $B_{(s)} \to S$, with $S$ denoting a
scalar meson, are investigated in the light-cone sum rules approach.
The numerical values are approximately twice as large as that
estimated in the light-front quark model and QCD sum rules approach.
Using these form factors, we present the analysis of the decay rates
for $B \to a_0(1450) l \bar{\nu}_l $, $B \to K^{\ast}_0(1430) l
\bar{l}$, $B_s \to K^{\ast}_0(1430) l \bar{\nu}_l$ and $B_s \to
f_0(1500) l \bar{l}$ with $l=e, \mu, \tau$. The results indicate
that magnitudes of $BR(\bar{B}_0 \to a_0(1450) l \bar{\nu}_l)$ and
$BR(B_s \to K^{\ast}_0(1430) l \bar{\nu}_l)$ can arrive at the order
of $10^{-4}$, which can be measured in the future experiments to
clarify the inner structure of scalar mesons. It is also observed
that $BR(B \to K^{\ast}_0(1430) \tau^+ \tau^-)$ and $BR(B_s \to
f_0(1500) \tau^+ \tau^-)$ are an order of magnitude smaller than the
corresponding channels of $e^+ e^-$ and $\mu^+ \mu^-$ final states
due to the heavily suppressed phase space . Moreover, the
longitudinal lepton polarization asymmetry for $B \to
K^{\ast}_0(1430) l \bar{l}$ and $B_s \to f_0(1500) l \bar{l}$ are
also investigated, whose values are close to $-1$ for the $e^+ e^-$
and $\mu^+ \mu^-$ pair except  the region close to the end points.
\end{abstract}

 \maketitle

\section{\protect\bigskip Introduction}

The inner structure of scalar mesons has been controversial for over
three decades, which  makes them one of the alluring issues in
contemporary particle physics. In particular, the existence of some
physical states, such as $f_0(1370)$, is still in dispute due to the
absence of convincing evidence \cite{Klempt:2007cp}. It is suggested
that the scalar mesons with the masses below and above $1
{\rm{GeV}}$ can be organized into two nonets in terms of their
spectrum \cite{Close, Cheng}. The flavor singlet $f_0(600)$ (or
$\sigma$), $f_0(980)$, the isodoublet $K^{\ast}_0(800)$ (or
$\kappa$), and the isovector $a_0(980)$ constitute the nonet below 1
GeV; while $f_0(1370)$, $f_0(1500)/f_0(1710)$, $K^{\ast}_0(1430)$
and $a_0(1450)$  form the other one near 1.5 GeV \cite{Close,
Cheng}. Up to now, there is no general agreement on the nature of
these states \cite{Spainer} due to the ambiguity existing in all the
available interpretations including conventional $q \bar{q}$ states
\cite{Cheng}, glueball, hybrid states, molecule states
\cite{Weinstein:1982gc, Weinstein:1983gd, Weinstein:1990gu} as well
as tetra-quark states \cite{Jaffe} and the superpositions of these
contents \cite{Amsler:1995td, Amsler:1995tu, Amsler:2002ey}. Among
all the scalar mesons, the $K^{\ast}_0(1430)$ is predominantly
viewed as $s \bar{u}$ or $s \bar{d}$ state in almost all the models.
Hence, it is justified to assume that $a_0(1450)$,
$K^{\ast}_0(1430)$ and $f_0(1370)$ being in the same nonent are
respectively the $u\bar{d}$, $u \bar{s}$ and ${u \bar{u}+ d \bar{d}
}$ states, based on the naive quark model, which is also the picture
of scalar mesons adopted in this paper.

Studies on the strong and electromagnetic decays of scalar decays
have been received extensive interests in the literature
\cite{StrohmeierPresicek:1999yv, Maiani:2004uc, Giacosa:2005qr,
Giacosa:2005zt}. Besides, the production properties of scalar mesons
in $\pi N$ scattering, $p \bar{p}$ annihilation, $\gamma \gamma$
formation and heavy meson decays can also serve as an ideal platform
to explore the underlying structures of scalar mesons as well as the
non-perturbative dynamics of QCD. Thanks to the progress of
accelerator and detector techniques, both Belle and BaBar have
observed strong indications of scalar mesons within a broad spectrum
between $1.0$ and $1.5$ GeV in $B$ meson decays \cite{Abe:2002av}.
In this work, we will focus on the semi-leptonic weak production of
scalars in $B_{(s)}$ decays, which are relatively clean compared to
the hadronic decays from the theoretical viewpoint. Moreover,
semileptonic decays of $B_{(s)}$ mesons  are also of great
importance to determine the quark-flavor mixing matrix
--- the Cabibbo-Kobayashi-Maskawa (CKM) matrix \cite{CKM 1, CKM 2}
and testing its unitarity under the requirement of the standard
model (SM).

The main job of investigating the semi-leptonic decays of $B_{(s)}$
to the scalar mesons ($S$) is to properly evaluate the hadronic
matrix elements for $B_{(s)} \to S$, namely the transition form
factors, which are governed by the non-perturbative QCD dynamics.
Several methods exist in the literature to deal with this problem,
such as simple quark model \cite{quark model}, light-front approach
\cite{light front QCD 1,light front QCD 2,light front QCD 3}, QCD
sum rules (QCDSR) \cite{QCDSR 1,QCDSR 2}, light-cone QCD sum rules
(LCSR) \cite{LCQCDSR 1,LCQCDSR 2,LCQCDSR 3}, perturbative QCD
factorization approach \cite{PQCD 1,PQCD 2,PQCD 3}. The QCD sum
rules approach, which is a fully relativistic approach and well
rooted in quantum field theory, has made a tremendous success;
however, short distance expansion fails in non-perturbative
condensate when applying the three-point sum rules to the
computations of form factors in the large momentum transfer or large
mass limit of heavy meson decays. As a marriage of standard QCDSR
technique \cite{QCDSR 1,QCDSR 2} and theory of hard exclusive
process \cite{hard exclusive process 1,hard exclusive process 2,hard
exclusive process 3,hard exclusive process 4,hard exclusive process
5,hard exclusive process 6,hard exclusive process 7,hard exclusive
process 8}, LCSR cure the problem of QCDSR applying to the large
momentum transfer by performing the operator product expansion(OPE)
in terms of twist of the revelent operators rather than their
dimension \cite{braun talk}.  An important advantage of light-cone
QCD sum rules is that it allows a systematic inclusion of both hard
scattering effects and the soft contributions \cite{perspective of
QCDSR}. Phenomenologically, LCSR has been widely applied to
investigate the semi-leptonic decays  of heavy hadrons
\cite{Ball:1998tj, Khodjamirian:2000ds, Duplancic:2008ix,
Wang:2007fs}, radiative hadronic decays \cite{Ali:1993vd,
Aliev:1995wi,Wang:2008sm}, non-leptonic two body decays of $B$ meson
\cite{Khodjamirian:2000mi,Khodjamirian:2002pk,Khodjamirian:2003eq,Khodjamirian:2005wn}
and strong coupling constants \cite{Belyaev:1994zk}.

In the present work, we would like to adopt LCSR approach to study
the rare decay of $B_{(s)} \to S$. The essential inputs in the
light-cone QCD sum rules is the hadronic distribution amplitudes
other than vacuum condensates in the QCD sum rules.  It is known
that LCDAs are non-perturbative functions, which describes the
hadronic structure in rare parton configurations with a fixed number
of Fock components at small transverse separation in the infinite
momentum frame. In an attempt to accommodate the experimental data,
there have been continuous interests concentrating on the research
of pre-asymptotic corrections to the distribution amplitudes of
hadrons in the exclusive reactions over two decades. In particular,
the leading twist and twist-3 distribution amplitudes of scalar
mesons have been worked out in \cite{Cheng,y.m. wang} based on the
QCD sum rules and conformal symmetry hidden in the QCD Lagrangian
and we will use these amplitudes  in this paper.

The paper is organized as follows: In section II we present the
effective Hamiltonian responsible for the  $b \to u,s$ transitions
in the standard model, where the parameterizations of hadronic
matrix elements are also collected here. Based on the trace formulae
and equation of motion, we also derive the relations among form
factors $f_{+}(q^2)$, $f_{-}(q^2)$ and $f_{T}(q^2)$ in the large
recoil and heavy quark limit. The Gegenauber moments of twist-2 and
twist-3 distribution amplitudes obtained in the QCD sum rules are
collected in section III. Then the sum rules for the various form
factors on the light-cone are derived  in section IV with the
standard correlation function to the leading Fork state.  After
grouping the input parameters, the numerical computations of form
factors in light-cone QCD sum rules are performed in section IV.
Subsequently, we apply these form factors to analyze the decay rates
of $\bar{B}_0 \to a_0(1450) l \bar{\nu}_l$, $\bar{B}_0 \to
K^{\ast}_0(1430) l \bar{l}$, $B_s \to K^{\ast}_0(1430) l
\bar{\nu}_l$ and $B_s \to f_0(1500) l \bar{l}$ as well as the
longitudinal lepton polarization asymmetry for the modes induced by
the flavor-changing neutral current, where a brief analysis on
comparisons with the results obtained  in the light-front quark
model and QCD sum rules are also included in this section. The last
section is devoted to the conclusions.

\section{Effective Hamiltonian and parameterizations of matrix element}

\subsection{Effective Hamiltonian for $b \to u,s$ transition}
Integrating out the particles including top quark, $W^{\pm}$ and $Z$
bosons above scale $\mu=O(m_b)$ , we  arrive at the effective
Hamiltonian responsible for the  $b \to u$ transition
\begin{eqnarray}
 \mathcal{H}_{eff}(b \to u l\bar \nu_l)={G_{F} \over \sqrt{2}}V_{ub}
 \bar{u}\gamma_{\mu}(1-\gamma_5)b \,
 \bar{l}\gamma^{\mu}(1-\gamma_5)\nu_l +h.c. \, ,
 \label{effectiveH b to u}
\end{eqnarray}
where $V_{ub}$ is the corresponding Cabbibo-Kobayashi-Maskawa (CKM)
matrix element and $l=\left( e,\mu ,\tau \right) $ .

Similarly, the effective Hamiltonian revelent to the flavor-changing
neutral current (FCNC) transition $b\rightarrow s$ can be derived as
\begin{eqnarray}
\mathcal{H}_{eff}(b &\rightarrow &sl\bar{l})={\frac{G_{F}}{\sqrt{2}}}%
V_{tb}V_{ts}^{*}[C_{9}^{eff}\left( \mu\right) \bar{s}\gamma _{\mu
}(1-\gamma _{5})b\,\bar{l}\gamma ^{\mu }(1-\gamma
_{5})l+C_{10}\bar{s}\gamma _{\mu
}(1-\gamma _{5})b\,\bar{l}\gamma ^{\mu }\gamma _{5})l  \nonumber \\
&&-\frac{2m_{b}C_{7}\left( \mu\right) }{q^{2}}\sigma _{\mu \nu
}(1-\gamma _{5})q^{\nu }b\,\bar{l}\gamma ^{\mu }l]+h.c. \,,
\label{effectiveH b to s}
\end{eqnarray}%
where we have neglected the terms proportional to $V_{ub}V_{us}^{*}$
on account of $|V_{ub}V_{us}^{*}/V_{tb}V_{ts}^{*}|<0.02$ . The
$C_{i}$ involved in Eq. (\ref {effectiveH b to s}) are the Wilson
coefficients and their particular expressions are given in Ref.
\cite{Buchalla}. We should emphasize that the Wilson coefficient
$C_{10}$ does not renormalize under QCD corrections and hence is
independent on the energy scale $\mu \simeq O(m_b)$, since the
operator $O_{10}={\bar{s}} \gamma _{\mu }(1-\gamma _{5})
b{\bar{l}}\gamma ^{\mu }\gamma _{5} l$ can not be induced by the
insertion of four-quark operators due to the absence of $Z$ boson in
the effective theory. Moreover, the above quark decay amplitude can
also receive additional contributions from the matrix element of
four-quark operators, which are usually absorbed into the effective
Wilson coefficient $C_9^{eff}(\mu)$. To be more specific, we can
decompose $C_9^{eff}(\mu)$ into the following three parts \cite{b to
s in theory 3,b to s in theory 4,b to s in theory 5,b to s in theory
6,b to s in theory 7,b to s in theory 8,b to s in theory 9}
\begin{eqnarray}
C_9^{eff}(\mu) = C_9(\mu)+Y_{SD}(z,s')+Y_{LD}(z,s'),
\end{eqnarray}
where the parameters $z$ and $s'$ are defined as $z=m_c/m_b, \,\,\,
s'=q^2/m_b^2$. $Y_{SD}(z,s')$ describes the short-distance
contributions from four-quark operators far away form the $c\bar{c}$
resonance regions, which can be calculated reliably in perturbative
theory. The long-distance contributions $Y_{LD}(z,s')$ from
four-quark operators near the $c\bar{c}$ resonance cannot be
calculated from first principles of QCD and are usually
parameterized in the form of a phenomenological Breit-Wigner
formula, which will be neglected in this work due to the absence of
experimental data on $B_{(s)} \to J/\psi S$.

The manifest expressions for $Y_{SD}(z,s')$ can be written as
\cite{Buchalla}
\begin{eqnarray}
Y_{SD}(z,s')&=&h(z,s')(3C_1(\mu)+C_2(\mu)+3C_3(\mu)+C_4(\mu)+3C_5(\mu)+C_6(\mu))\nonumber\\
&&-\frac{1}{2}h(1,s')(4C_3(\mu)+4C_4(\mu)+3C_5(\mu)
+C_6(\mu))\nonumber\\
&&-\frac{1}{2}h(0,s')(C_3(\mu)+3C_4(\mu))+{2 \over
9}(3C_3(\mu)+C_4(\mu)+3C_5(\mu) +C_6(\mu)),
\end{eqnarray}
with
\begin{eqnarray}
h(z,s') &=& -{8\over 9}{\rm {ln}}z+{8\over 27}+{4\over 9}x-{2\over
9}(2+x)|1-x|^{1/2}
 \left\{
\begin{array}{l}
\ln \left| \frac{\sqrt{1-x}+1}{\sqrt{1-x}-1}\right| -i\pi \quad {\rm {for}}%
{ {\ }x\equiv 4z^{2}/s^{\prime }<1} \\
2\arctan \frac{1}{\sqrt{x-1}}\qquad {\rm {for}}{ {\ }x\equiv
4z^{2}/s^{\prime }>1}
\end{array}
\right., \nonumber\\
h(0,s^{\prime}) &=& {8 \over 27}-{8 \over 9} {\rm ln}{m_b \over \mu}
-{4 \over 9} {\rm {ln}}s^{\prime} +{4 \over 9}i \pi \,\, .
\end{eqnarray}

Besides, the non-factorizable effects from the charm quark loop can
bring about further corrections to the radiative $b \to s \gamma$
transition, which can be absorbed into the effective Wilson
coefficient $C_7^{eff}$ as usual \cite{b to s 1, b to s 2, b to s
3,NF charm loop}. Specifically, the Wilson coefficient $C^{eff}_{7}$
is given by \cite{c.q. geng 4}
\begin{eqnarray}
C_7^{eff}(\mu) = C_7(\mu)+C'_{b\to s\gamma}(\mu),
\end{eqnarray}
with
\begin{eqnarray}
C'_{b\rightarrow s \gamma}(\mu) &=& i\alpha_s[{2\over
9}\eta^{14/23}(G_1(x_t)-0.1687)-0.03C_2(\mu)], \\
G_1(x)  &=& {x(x^2-5x-2)\over 8(x-1)^3}+{3x^2 {\rm{ln}}^2x\over
4(x-1)^4},
\end{eqnarray}
where $\eta=\alpha_s(m_W)/\alpha_s(\mu)$, $x_t=m_t^2/m_W^2$,
$C'_{b\rightarrow s\gamma}$ is the absorptive part for the $b \to s
c \bar{c} \to s \gamma$ rescattering and we have dropped out the
tiny contributions proportion to CKM sector $V_{ub}V_{us}^{\ast}$.

\subsection{Parameterizations of hadronic matrix element}

With the free quark decay amplitude available, we can proceed to
calculate the decay amplitudes for semi-leptonic decays of $B_{q'}
\to S$ at hadronic level, which can be obtained by sandwiching the
free quark amplitudes between the initial and final meson states.
Consequently, the following two hadronic matrix elements
\begin{eqnarray}
\langle S(p)|\bar{s}\gamma _{\mu }\gamma _{5}b|B_{q^{\prime
}}(p+q)\rangle, \,\,\,   \langle S(p)|\bar{s}\sigma _{\mu \nu
}\gamma _{5}q^{\nu }b|B_{q^{\prime }}(p+q)\rangle
\end{eqnarray}
need to be computed as can be observed from Eqs. (\ref{effectiveH b
to u}) and (\ref{effectiveH b to s}). The contributions from vector
and tensor types of transitions vanish due to parity conservations
which is the property of strong interactions. Generally, the above
two matrix elements can be parameterized in terms of a series of
form factors as
\begin{eqnarray}
\langle S(p)|\bar{s}\gamma _{\mu }\gamma
_{5}b|B_{q^{\prime }}(p+q)\rangle
&=&-i[f_{+}(q^{2})p_{\mu }+f_{-}(q^{2})q_{\mu }],
\label{axial form factor} \\
\langle S(p)|\bar{s}\sigma _{\mu \nu }\gamma _{5}q^{\nu
}b|B_{q^{\prime }}(p+q)\rangle  &=&-\frac{1}{m_{B}+m_{S}}\left[
\left( 2p+q\right) _{\mu }q^{2}-\left( m_{B}^{2}-m_{S}^{2}\right)
q_{\mu }\right] f_{T}\left( q^{2}\right). \label{tensor form factor}
\end{eqnarray}

Utilizing the covariant trace formalism introduced in
\cite{Falk:1990yz}, the form factors at large recoil should satisfy
the following relations
\begin{eqnarray}
f_{+}(q^2)={2 m_B \over m_B+m_S} f_{T}(q^2), \qquad  f_{-}(q^2)=0,
\label{relations of  form factor}
\end{eqnarray}
where the corrections due to hard gluon exchange are neglected
\cite{Beneke:2000wa}. We can also derive the relation \footnote{This
relation has been derived in Ref. \cite{Aliev:2007rq},  where the
convention $\sigma_{\mu \nu}=-{i \over 2} (\gamma_{\mu} \gamma_{\nu}
- \gamma_{\nu} \gamma_{\mu})$ is adopted . Therefore, the relations
between $f_{-}(q^2)$ and $f_{T}(q^2)$ there differ from that given
in this paper with a minus sign.} between $f_{-}(q^2)$ and
$f^{T}(q^2)$ as
\begin{eqnarray}
f_{T}(q^2)=-{m_b -m_{q_2} \over m_{B} -m_{S}} f_{+}(q^2),
\end{eqnarray}
with the help of the equation of motion in the  heavy quark limit
\cite{Gilani:2003hf}.

\section{Distribution amplitudes of scalar mesons}
\label{Distribution amplitudes of scalar mesons}

In this section, we would like to collect some revelent information
on the distribution amplitudes for scalar mesons, which are the
essential ingredients in the sum rules for the $B \to S$ transition
form factors.

Up to the leading Fock states, the light-cone distributions of
scalar mesons made up of $q_2 \bar{q}_1$ can be defined as
\cite{Cheng, y.m. wang}:
\begin{eqnarray}
\left\langle S\left( p\right) \left| \bar{q}_{2}\left( x\right)
\gamma _{\mu }q_{1}\left( y\right) \right| 0\right\rangle
&=&p_{u}\int_{0}^{1}due^{i\left( up\cdot x+\bar{u}p.y\right) }\Phi
_{S}\left( u,\mu \right),  \nonumber \\
\left\langle S\left( p\right) \left| \bar{q}_{2}\left( x\right)
q_{1}\left( y\right) \right| 0\right\rangle
&=&m_{S}\int_{0}^{1}due^{i\left( up\cdot x+\bar{u}p.y\right) }\Phi
_{S}^{s}\left( u,\mu \right),  \label{DAs}
\\
\left\langle S\left( p\right) \left| \bar{q}_{2}\left( x\right)
\sigma _{\mu \nu }q_{1}\left( y\right) \right| 0\right\rangle
&=&-m_{S}\left(
p_{\mu }z_{\nu }-p_{\nu }z_{\mu }\right) \int_{0}^{1}due^{i\left( up\cdot x+%
\bar{u}p.y\right) }\Phi _{S}^{\sigma }\left( u,\mu \right),
\nonumber
\end{eqnarray}
where $z=x-y$, $m_S$ is the mass of corresponding scalar meson,
$\bar{u}=1-u$ and $u$ is the momentum fraction carried by the quark
$q_{2}$ in the scalar meson. Here $\Phi _{S}\left( u,\mu \right)$ is
of twist-2, $\Phi _{S}^{s}\left( u,\mu \right) $ and  $\Phi
_{S}^{\sigma }\left( u,\mu \right)$ are of twist-3. $\Phi _{S}\left(
u,\mu \right)$ and ($\Phi _{S}^{s}\left( u,\mu \right)$, $\Phi
_{S}^{\sigma }\left( u,\mu \right)$) are anti-symmetric and
symmetric under the replacement of $u \to 1-u$ in the SU(3) limit
owing to the conservation of G parity. To be more specific, the
normalizations of $\Phi _{S}\left( u, \mu \right) $, $\Phi
_{S}^{s}\left( u, \mu \right) $ and $\Phi _{S}^{\sigma }\left( u,
\mu \right) $ are given by
\begin{equation}
\int_{0}^{1}du\Phi _{S}\left( u,\mu \right) =f_{S}\text{, } \,\,\,
\int_{0}^{1}du\Phi _{S}^{s}\left( u,\mu \right) =\int_{0}^{1}du\Phi
_{S}^{\sigma }\left( u,\mu \right) =\bar{f}_{S}.
\label{normalization}
\end{equation}
The vector current decay constant $f_{S}$ defined by
\begin{equation}
\left\langle S\left( p\right) \left| \bar{q}_{2}\gamma ^{\mu
}q_{1}\right| 0\right\rangle =f_{S}p^{\mu }
\end{equation}
should vanish in the SU(3) limit and can be related to the scalar
density decay constant $\bar{f}_{S}$ determined by
\begin{equation}
\left\langle S\left( p\right) \left| \bar{q}_{2}q_{1}\right|
0\right\rangle =m_{S}\bar{f}_{S}
\end{equation}
with the help of the equations of motion
\begin{equation}
\bar{f}_{S}=\mu _{S}f_{S}\text{, \ \ \ \ \ \ \ \ \ \ \ }\mu _{S}\text{\ =}%
\frac{m_{S}}{m_{2}\left( \mu \right) -m_{1}\left( \mu \right) },
\label{relation of decay constant}
\end{equation}
with $m_{1}$ and $m_{2}$ being the masses of quarks $q_{1}$ and
$q_{2}$ respectively. It should be emphasized that scalar density
meson decay constant $\bar{f}_{S}$ depends on the renormalization
scale $\mu$, whereas the the vector current decay constant $f_{S}$
does not renormalize under the QCD corrections due to the
conservation of vector current.

In view of the conformal symmetry hidden in the QCD Lagrangian, the
distribution amplitudes of scalar mesons $\Phi _{S}\left( u, \mu
\right) $, $\Phi _{S}^{s}\left( u, \mu \right) $ and $\Phi
_{S}^{\sigma }\left( u, \mu \right) $ can be expanded in the Hilbert
space by Jacobbi polynomials with increasing conformal spin as
\begin{eqnarray}
\Phi _{S}\left( u,\mu \right) &=&\bar{f}_{S}\left( \mu \right) 6u\bar{u}%
\left[ B_{0}\left( \mu \right) +\sum_{m=1}^{\infty }B_{m}\left( \mu
\right)
C_{m}^{3/2}\left( 2u-1\right) \right],  \nonumber \\
\Phi _{S}^{s}\left( u,\mu \right) &=&\bar{f}_{S}\left( \mu \right) \left[
1+\sum_{m=1}^{\infty }a_{m}\left( \mu \right) C_{m}^{1/2}\left( 2u-1\right) %
\right] , \label{Gegexpansion} \\
\Phi _{S}^{\sigma }\left( u,\mu \right) &=&\bar{f}_{S}\left( \mu \right) 6u%
\bar{u}\left[ 1+\sum_{m=1}^{\infty }b_{m}\left( \mu \right)
C_{m}^{3/2}\left( 2u-1\right) \right],  \nonumber
\end{eqnarray}
where Gegenbauer polynomial $C_{m}^{3/2}(x)$ can be considered as a
special type of Jacobbi polynomials $P_{m+2}^{1,1}(x) \sim
C_{m}^{3/2}(x)$. Combining Eqs.(\ref{normalization}), (\ref{relation
of decay constant}) and (\ref{Gegexpansion}), the zeroth Gegenbauer
moment $B_0(\mu)$ for twist-2 distribution amplitude $\Phi_S(u,\mu)$
is given by
\begin{eqnarray}
B_{0}=\text{\ }\mu _{S}^{-1}.
\end{eqnarray}

\begin{table}[tb]
\caption{Decay constants and  Gegenbauer moments for the twist-2
distribution amplitude $\Phi_{S}$ of scalar mesons at the scale
$\mu=1 {\rm{GeV}}$ \cite{Cheng}. }
\begin{center}
\begin{tabular}{c c c c}
\hline\hline
 state &  $\bar{f}${(\rm{MeV})}& $B_1$ & $B_3$ \\
\hline $a_0(1450)$ &$460 \pm 50$&
$-0.58 \pm 0.12$ & $-0.49 \pm 0.15$    \\
\hline$K^*_0(1430)$&$445 \pm 50$& $-0.57 \pm 0.13$ & $-0.42 \pm 0.22$ \\
\hline$f_0(1500)$ & $490 \pm 50$& $-0.48 \pm 0.11$& $-0.37 \pm 0.20$  \\
\hline \hline
\end{tabular}
\end{center}\label{results1}
\end{table}

\begin{table}[tb]
\caption{Gegenbauer moments for the twist-3 distribution amplitudes
$\Phi_{S}^{s}$ and $\Phi_{S}^{\sigma}$ of scalar mesons at the scale
$\mu=1 {\rm{GeV}}$ \cite{y.m. wang}.  }
\begin{center}
\begin{tabular}{c c c c c c c}
\hline \hline  state &  $a_1 (\times 10^{-2})$ & $a_2$ & $a_4$ &  $b_1 (\times 10^{-2})$ & $b_2$ & $b_4$\\
\hline $a_0$ &0 & $-0.33 \sim -0.18 $   & $-0.11 \sim 0.39$ & 0 & $0 \sim 0.058$ & $0.070 \sim 0.20$ \\
\hline$K^*_0$& $1.8 \sim 4.2$ & $-0.33 \sim -0.025$ & --- & $3.7 \sim 5.5$& $0 \sim 0.15$ & ---  \\
\hline$f_0$& 0 & $-0.33 \sim -0.18$ & $0.28 \sim 0.79$  & 0 & $-0.15 \sim -0.088$& $0.044 \sim 0.16$\\
\hline \hline
\end{tabular}
\end{center}\label{results2}
\end{table}

Moreover, decay constants of scalar mesons and  various Gengauber
moments $B_m$, $a_m$ and $b_m$ for both twist-2 and twist-3 LCDAs
have been computed in Refs. \cite{Cheng, y.m. wang} based on QCD sum
rules approach, which are collected in Table \ref{results1} and
\ref{results2}.

\section{Light Cone Sum Rules for Form Factors}

With the LCDAs of scalar mesons available, we are now in a position
to derive the sum rules of transition form factors which are
responsible for $B_{(s)} \to S$ decays. The basics object in LCSR
approach is the correlation function in which one of the hadron is
represented by the interpolating current with proper quantum number,
such as
 spin, isospin, (charge) parity  and so on; and the
other is described by its vector state manifestly. Information on
the hadronic transition form factor can be extracted by matching the
Green function calculated in two different representations, i.e.,
phenomenological and theoretical forms, with the help of  dispersion
relation under the assumption of quark-hadron duality.

\subsection{Light-cone sum rules for the form factors $f_{+}( q^{2}) $
 and $f_{-}( q^{2})$}

Following the standard procedure of sum rules, we consider the
correlation function associating with the form factors $f_{+}\left(
q^{2}\right)$ and $f_{-}\left( q^{2}\right)$ determined by the
matrix element
\begin{eqnarray}
\Pi _{\mu }\left( p,q\right) &=&-\int d^{4}xe^{iqx}\left\langle
S\left( p\right) \left| T\left\{ {j}_{2 \mu}(x), j_1(0)\right\}
\right| 0\right\rangle , \label{correlator1}
\end{eqnarray}%
where the current ${j}_{2 \mu}(x)=\bar{q}_{2}\left( x\right) \gamma
_{\mu }\gamma _{5} b\left( x\right)$ describes the weak transition
of $b$ to $q_2$ and $j_1(0)=b\left( 0\right) i\gamma _{5}q_1\left(
0\right)$ represents the  $B_{q_1}$ channel.  In addition, the
vacuum-to-meson matrix element for the interpolating current can be
given by
\begin{equation}
\left\langle B_{q_1}\left| \bar{b}i \gamma_5
q\right| 0\right\rangle =\frac{m_{B_{q_1}}^{2}}{%
m_{b}+m_{q_1}} f_{B_{q_1}}.  \label{decay constant of B meson}
\end{equation}
Inserting the complete set of states between the currents in Eq.
(\ref{correlator1}) with the same quantum numbers as $B_{q_1}$, we
can arrive at the hadronic representation of the correlator
(\ref{correlator1}):
\begin{eqnarray}
\Pi _{\mu }\left( p,q\right) &=& i {\frac{\langle S\left( p\right)
\left| \bar{q}_{2}\left( 0\right) \gamma _{\mu }\gamma _{5} b\left(
0 \right) \right|B_{q_1}(p+q)\rangle \langle
B_{q_1}(p+q)|\bar{b}\left( 0\right)
i\gamma_{5}q_{1}\left( 0\right) |0\rangle }{m_{B_{q_1}}^{2}-(p+q)^{2}}} \nonumber \\
&&+\sum_{h}i {\frac{\langle S\left( p\right) \left|
\bar{q}_{2}\left( 0\right) \gamma _{\mu }\gamma _{5} b\left( 0
\right) \right|h(p+q)\rangle \langle h(p+q)|\bar{b}\left( 0\right)
i\gamma_{5}q_{1}\left( 0\right) |0\rangle }{m_{h}^{2}-(p+q)^{2}}},
\label{hadronic level for correlator 1}
\end{eqnarray}
where we have separated the  contributions from the ground state and
higher states corresponding to the $B_{q_1}$ meson channel.
Combining the Eqs. (\ref{axial form factor}) , (\ref{decay constant
of B meson}) and (\ref{hadronic level for correlator 1}), the
phenomenological representations of correlation function
(\ref{correlator1}) can be derived as
\begin{eqnarray}
\Pi _{\mu }\left( p,q\right)
&=&\frac{m_{B_{q_1}}^{2}f_{B_{q_1}}}{\left( m_{b}+m_{q_1}\right) [
m_{B_{q_1}}^{2}-\left( p+q\right) ^{2} ] }
[f_{+}(q^2)p_{\mu}+f_{-}(q^2)q_{\mu}] +\int_{s_0^{B_{q_1}}}^{\infty}
ds { \rho^h_{+}(s,q^2) p_{\mu} +\rho^h_{-}(s,q^2) q_{\mu} \over
s-(p+q)^2}, \label{results for hadronic level for correlator 1}
\end{eqnarray}
where we have expressed the contributions from higher states of the
$B_{q_1}$ channel in the form of dispersion integral with
$s_0^{B_{q_1}}$ being the threshold parameter corresponding to the
$B_{q_1}$ channel.

On the theoretical side,  the correlation function
(\ref{correlator1}) can also be computed in the perturbative theory
with the help of OPE technique at the deep Euclidean region $p^2,
\,\, q^2=-Q^2 \ll 0$:
\begin{eqnarray}
\Pi_{\mu}(p,q)&=& \Pi_{+}^{QCD}(q^2,(p+q)^2) p_{\mu}
+\Pi_{-}^{QCD}(q^2,(p+q)^2) q_{\mu} \nonumber \\
&=&\int_{(m_b+m_{q_1})^2}^{\infty} ds {1 \over \pi} {{\rm{Im}}\,
\Pi_{+}^{QCD}(s,q^2) \over s-(p+q)^2} p_{\mu}
+\int_{(m_b+m_{q_1})^2}^{\infty} ds {1 \over \pi} { {\rm{Im}} \,
\Pi_{-}^{QCD}(s,q^2) \over s-(p+q)^2} q_{\mu}.
\end{eqnarray}
Making use of  the quark-hadron duality assumption
\begin{eqnarray}
\rho^h_{i}(s,q^2)={1 \over \pi}  {\rm{Im}}\, \Pi_{i}^{QCD}(s,q^2)
\Theta (s-s_0^h),
\end{eqnarray}
with $i=``+,-"$ and performing the Borel transformation
\begin{eqnarray}
\hat{\mathcal{B}}_{M^2}=\lim_{\stackrel{-(p+q)^2,n \to
\infty}{-(p+q)^2/n=M^2}} \frac{(-(p+q)^2)^{(n+1)}}{n!}\left(
\frac{d}{d(p+q)^2}\right)^n,
\end{eqnarray}
with variable $(p+q)^2$ to both two representations of the
correlation function, we can finally derive the sum rules for the
form factors
\begin{eqnarray}
f_i(q^2)={m_b+m_{q_1} \over \pi f_{B_{q_1}}m_{B_{q_1}}^2}
\int_{(m_b+m_{q_1})^2}^{s_0^{B_{q_1}}} {\rm{Im}} \,
\Pi_{i}^{QCD}(s,q^2) {\rm{exp}}\bigg({m_{B_{q_1}}^2-s \over
M^2}\bigg) ds. \label{formal sum rules for axial-vector}
\end{eqnarray}

To the leading order of $\alpha_s$, the correlation function can be
calculated by contracting the bottom quark field in Eq.
(\ref{correlator1}) and inserting the free $b$ quark propagator
\begin{eqnarray}
\Pi _{\mu }\left( p,q\right) &=&i\int d^{4}x \int \frac{d^{4}k}{%
\left( 2\pi \right) ^{4}}\frac{e^{i\left( q-k\right) x}}{m_{b}^{2}-k^{2}}%
\left\langle S\left( p\right) \left| \bar{q}_{2}\left( x\right)
\gamma _{\mu }\gamma _{5} \left( \not \!{k}+m_{b}\right) i \gamma
_{5}q_1\left( 0\right) \right| 0\right\rangle \label{theoretical
correlator 1},
\end{eqnarray}
which can be represented by Fig. (\ref{correlation function at quark
level}) intuitively.
\begin{figure}[tb]
\begin{center}
\begin{tabular}{ccc}
\includegraphics[scale=0.6]{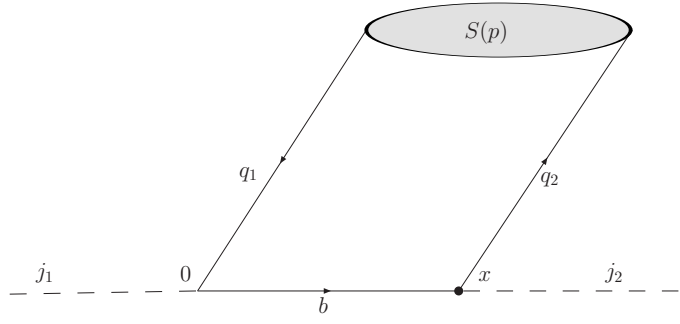}
\vspace{0 cm}
\end{tabular}
\caption{The tree level contributions to the correlation function
Eq. (\ref{correlator1}), where the current $j_{1}(0)$ represents the
$B_{q_1}$ channel and the current $j_{2}(x)$ describes  the $b \to
q_2$ transition.} \label{correlation function at quark level}
\end{center}
\end{figure}
It should be pointed out that the full quark propagator also
receives corrections from the background field \cite{asymptonic
forms 1,Khodjamirian:1998ji} and can be written as
\begin{eqnarray}
\langle 0| T \{{b_i(x) \bar{b}_j(0)}\}| 0\rangle &=& \delta_{ij}\int
{d^4 k \over (2 \pi)^4} e^{-i kx}{i \over \not \! k -m_b} -i g \int
{d^4 k \over (2 \pi)^4} e^{-i kx} \int_0^1 dv  [{1 \over 2} {\not k
+m_b \over (m_b^2 -k^2)^2} G^{\mu \nu}_{ij}(v x)\sigma_{\mu \nu }\nonumber \\
&& +{1 \over m_b^2-k^2}v x_{\mu} G^{\mu \nu}(v x)\gamma_{\nu}],
\end{eqnarray}
where the first term is the free-quark propagator and $G^{\mu
\nu}_{i j}=G_{\mu \nu}^{a} T^a_{ij}$ with ${\mbox{Tr}}[T^a T^b]={1
\over 2}\delta^{ab}$. Substituting the second term  proportional to
the gluon field strength into the correlation function can result in
the distribution amplitudes corresponding to the higher Fock states
of scalar mesons. It is expected that such corrections associating
with the LCDAs of higher Fock states do not play any significant
role in the sum rules for transition form factors \cite{higher Fock
state}, and  so can be safely neglected.

Substituting Eq.~(\ref{DAs}) into Eq.~(\ref{theoretical correlator
1}) and performing the the integral in the coordinate space, we can
achieve the correlation function in the momentum representation at
the quark level as
\begin{eqnarray}
\Pi _{\mu }\left( p,q\right) &=&p_{\mu }\int_{0}^{1}du%
\frac{1}{m_{b}^{2}-\left( q+up\right) ^{2}}\left\{ -m_{b}\Phi
_{S}\left( u\right) +um_{S}\Phi _{S}^{s}\left( u\right)
+\frac{1}{6}m_{S}\Phi
_{S}^{\sigma }\left( u\right) \left[ 2+\frac{m_{b}^{2}-u^{2}p^{2}+q^{2}}{%
m_{b}^{2}-\left( q+up\right) ^{2}}\right] \right\}  \nonumber \\
&&+q_{\mu }\int_{0}^{1}du\frac{1}{m_{b}^{2}-\left( q+up\right)
^{2}}\left\{ m_{S}\Phi _{S}^{s}\left( u\right) +\frac{m_{S}}{6u}\Phi
_{S}^{\sigma }\left( u\right) \left[
1-\frac{m_{b}^{2}+u^{2}p^{2}-q^{2}}{m_{b}^{2}-\left( q+up\right)
^{2}}\right] \right\} \nonumber \\
&\equiv& \Pi_{+}^{QCD}(q^2,(p+q)^2) p_{\mu}
+\Pi_{-}^{QCD}(q^2,(p+q)^2) q_{\mu} .  \label{results of theoretical
correlator 1}
\end{eqnarray}%

Combining the Eqs. (\ref{formal sum rules for axial-vector}) and
(\ref{results of theoretical correlator 1}), we can finally arrive
at the sum rules for form factors $f_{+}(q^2)$ and $f_{-}(q^2)$ as
below
\begin{eqnarray}
f_{+}\left( q^{2}\right) &=&\frac{\left( m_{b}+m_{q_1}\right) }{
m_{B_{q_1}}^{2}f_{B_{q_1}}}\exp \left( \frac{m_{B}^{2}}{M^2} \right)
\bigg \{ \int_{u_{0}}^{1}\frac{du}{u}\exp \left[
-\frac{m_{b}^{2}+u\bar{u}p^{2}-\bar{u}q^{2}}{uM^2}\right] \times
\bigg[ \bigg(-m_{b}\Phi _{S}\left( u\right) +m_{S}\left( u\Phi
_{S}^{s}\left( u\right) +\frac{1}{3}\Phi
_{S}^{\sigma }\left( u\right) \right) \bigg)\nonumber \\
&&\hspace{3.0 cm}+\frac{1}{uM^2}\frac{m_{S}}{6}\Phi _{S}^{\sigma
}\left( u\right) \left( m_{b}^{2}+u^{2}p^{2}+q^{2}\right) \bigg]
+\frac{m_{S}}{6}\Phi _{S}^{\sigma }\left( u_{0}\right) \exp \left( -\frac{%
s_{0}}{M^2}\right) \frac{m_{b}^{2}-u_{0}p^{2}+q^{2}}{
m_{b}^{2}+u_{0}^{2}p^{2}-q^{2}} \bigg \}, \label{fplus}
\\
f_{-}\left( q^{2}\right) &=&\frac{\left( m_{b}+m_{q_1}\right) }{%
m_{B_{q_1}}^{2}f_{B_{q_1}}}\exp \left( \frac{m_{B}^{2}}{M^2} \right)
\bigg \{ \int_{u_{0}}^{1}\frac{du}{u}\exp \left[
-\frac{m_{b}^{2}+u\bar{u}p^{2}-\bar{u}q^{2}}{uM^2}\right] \times
\bigg[ \bigg( m_{S}\left( \Phi _{S}^{s}\left( u\right)
+\frac{1}{6u}\Phi _{S}^{\sigma }\left(
u\right) \right) \bigg) \nonumber \\
&&\hspace{3.0 cm}-\frac{1}{u^{2}M^2}\frac{m_{S}}{6}\Phi _{S}^{\sigma
}\left( u\right) \left( m_{b}^{2}+u^{2}p^{2}-q^{2}\right)\bigg]
-\frac{m_{S}}{6u_{0}}\Phi _{S}^{\sigma }\left( u_{0}\right) \exp
\left(-\frac{s_{0}}{M^2}\right)\bigg \}, \label{fminus}
\end{eqnarray}
with
\begin{equation}
u_{0}={\frac{-(s_{0}-q^{2}-p^{2})+\sqrt{%
(s_{0}-q^{2}-p^{2})^{2}+4p^{2}(m_{b}^{2}-q^{2})}}{2p^{2}}.}
\label{unot}
\end{equation}
As can be observed from the sum rules (\ref{fplus}) and
(\ref{fminus}), both twist-2 and twist-2 distribution amplitudes of
scalar mesons can contribute to the form factor $f_{+}(q^2)$,
whereas the other one $f_{-}(q^2)$ can only receive the
contributions form twist-3 LCDAs and should be heavily suppressed in
the large recoil region, which is also in agreement with the
relations (\ref{relations of  form factor}) presented in
\cite{Beneke:2000wa}.

\subsection{Light-cone sum rules for the form factor $f_{T} ( q^{2})$}

As for the form factor  $f_{T}\left( q^{2}\right)$ involved in $b
\to s$ transition, we start with the following correlation function
\begin{eqnarray}
\tilde{\Pi} _{\mu }\left( p,q\right) &=&-\int
d^{4}xe^{iqx}\left\langle S\left( p\right) \left| T\left\{
{\tilde{j}_{2 \mu}}(x), j_1(0)\right\} \right| 0\right\rangle
\label{correlator 2}
\end{eqnarray}%
where the current ${\tilde{j}_{2 \mu}}(x)$ is given by
\begin{eqnarray}
{\tilde{j}_{2 \mu}}(x)= \bar{ q}_2(x) \sigma_{\mu \nu}a^{\nu}
\gamma_5 b(x)\,\,.
\end{eqnarray}

One can write the phenomenological representation of the correlation
function at the hadronic level simply by repeating the procedure
given above as
\begin{eqnarray}
\tilde{\Pi} _{\mu }\left( p,q\right)
&=&\frac{m_{B_{q_1}}^{2}f_{B_{q_1}}}{\left( m_{b}+m_{q_1}\right) [
m_{B_{q_1}}^{2}-\left( p+q\right) ^{2} ] (m_B+m_S)}
[(2p+q)_{\mu}q^2-q_{\mu}(m_{B}^2-m_{S}^2)] f_{T}(q^2) \nonumber \\
&&+\int_{s_0^{B_{q_1}}}^{\infty} ds { 1 \over s-(p+q)^2}
[-p_{\mu}q^2+q_{\mu} (p \cdot q)] \rho^h_{T}(s,q^2). \label{results
for hadronic level for correlator 2}
\end{eqnarray}
On the other hand, the correlation function at the quark level can
be calculated in the framework of perturbative theory to the leading
order of $\alpha_s$ as
\begin{equation}
\tilde{\Pi} _{\mu }\left( p,q\right) =[-p_{\mu}q^2+q_{\mu} (p \cdot
q)]  \int_{0}^{1}du\frac{1}{m_{b}^{2}-\left( q+up\right)
^{2}}\left\{ \Phi _{S}\left( u\right)
-\frac{m_{b}m_{S}}{3}\frac{\Phi _{S}^{\sigma }\left( u\right)
}{m_{b}^{2}-\left( q+up\right) ^{2}}\right\} .  \label{results for
theoretical level for correlator 2}
\end{equation}

Matching the correlation function obtained in the two different
representations and performing the Borel transformation with respect
to the variable $(p+q)^2$, we can achieve the sum rules for the form
factor $f_T(q^2)$

\begin{eqnarray}
f_{T}\left( q^{2}\right) &=&\frac{\left( m_{b}+m_{q_1}\right) (m_B+m_S)  }{%
m_{B_{q_1}}^{2}f_{B_{q_1}} }\exp \left( \frac{m_{B}^{2}}{M^2}%
\right)  \bigg \{ -\frac{1}{2}\int_{u_{0}}^{1}\frac{du}{u}\exp \left[ -\frac{m_{b}^{2}+u\bar{u}%
p^{2}-\bar{u}q^{2}}{uT}\right] \times \bigg[ \Phi _{S}\left(
u\right) -\frac{m_{B}m_{S}}{3uM^2} \Phi _{S}^{\sigma }\left(
u\right)  \bigg] \nonumber \\
&&\hspace{3.0 cm} +\frac{m_{b}m_{S}}{6}\Phi _{S}^{\sigma }\left( u_{0}\right) \exp \left( -%
\frac{s_{0}}{M^2}\right) \frac{1}{m_{b}^{2}+u_{0}^{2}p^{2}-q^{2}}
\bigg \}. \label{ftensor}
\end{eqnarray}

\section{Numerical analysis of transition form factors}

Now we are going to analyze the sum rules for the form factors
numerically.  Firstly, we collect the input parameters used in this
paper as below \cite{PDG, Kiselev, korner, bauer, ioffe,
Gray:2005ad, Penin:2001ux, Jamin:2001fw}:
\begin{equation}
\begin{array}{ll}
G_F=  1.166 \times 10^{-2} {\rm{GeV}^{-2}}, &
|V_{ub}|=3.96^{+0.09}_{-0.09} \times 10^{-3},
\\
|V_{tb}|=0.9991,  & |V_{ts}|=41.61^{+0.10}_{-0.80} \times 10^{-3},
\\
m_b=(4.68 \pm 0.03) {\rm{GeV}}, &m_s(1 {\rm{GeV}})=142 {\rm{MeV}},
\\
m_u(1 {\rm{GeV}})=2.8 {\rm{MeV}}, &  m_d(1 {\rm{GeV}})=6.8
{\rm{MeV}},
\\
m_{B_0}=5.279 {\rm{GeV}}, &  m_{B_s}=5.368 {\rm{GeV}},
\\
f_{B_0}=(0.19 \pm 0.02) {\rm{GeV}}, & f_{B_s}=(0.23 \pm 0.02)
{\rm{GeV}}. \label{inputs}
\end{array}
\end{equation}

It is noted that the input values for $f_B$ and $f_{B_s}$ are in
agreement with the unquenched lattice results~\cite{Gray:2005ad}
$f_B=0.216\pm\ 0.022 \mbox{ GeV}$ and $f_{B_s}=0.259 \pm\ 0.032
\mbox{  GeV}$, and with the results from the QCD sum
rules~\cite{Penin:2001ux,Jamin:2001fw}. The threshold parameter $s$
can be determined by the condition that the sum rules should take on
the best stability in the allowed Borel region. Besides, the values
of threshold parameter should be around the mass square of the
corresponding first excited state, hence they are also chosen the
same as that in the usual two-point QCD sum rules. The standard
value of the threshold in the $X$ channel is
${s_0}_{X}=(m_X+\Delta_X)^2$, where $\Delta_X$ is approximately
taken to be $0.6 \mathrm{GeV}$ in the literature \cite{dosch,
matheus, bracco, navarra} . To be more specific, we adopt the
threshold parameters $s_0^{B_0}=(35 \pm 2){\rm{GeV}}^2$ and
$s_0^{B_s}=(36 \pm 2){\rm{GeV}}^2$ corresponding to $B_0$ and $B_s$
channels respectively,  for the error estimate in the numerical
analysis.

With  all the parameters, we can proceed to compute the numerical
values of the form factors. In principle, the form factors
$f_{+}(q^2)$, $f_{-}(q^2)$  and $f_T(q^2)$ should not depend on the
Borel mass $M^2$ in a complete theory. However, as we truncate the
operator product expansion up to the leading conformal spin of
distribution amplitudes for scalar mesons in the leading Fock
configuration and keep the perturbative expansion in $\alpha_s$ to
leading order, a manifest dependence of the form factors on the
Borel parameter $M^2$ would emerge  in practice. Therefore, one
should look for a working ``window", where the results only mildly
vary with respect to the Borel mass, to make the truncation
reasonable and acceptable.

Firstly,  we concentrate on  the form factors at zero momentum
transfer.  As for the form factors $f_{+}(0)$ involved in $\bar{B}_0
\to a_0(1450) l \nu_l$, we require that the contributions from the
higher excited resonances and continuum states hold the fraction
less than 20 \% in the total sum rules and the value of $f_+(0)$
does not vary drastically within the selected region for the Borel
mass.  In view of these considerations, the Borel parameter $M^2$
should not be too large in order to ensure that the contributions
from the higher states are exponentially damped as can be observed
from Eqs. (\ref{fplus}), (\ref{fminus}) and (\ref{ftensor}) and the
global quark-hadron duality is satisfactory; on the other hand, the
number of Borel mass also could not be too small for the sake of
validity of OPE near the light-cone for the correlation function in
deep Euclidean region, since the contributions of higher twist
distribution amplitudes amount the higher order of ${1 / M^2}$ to
the perturbative part. Subsequently, we  indeed find the Borel
platform $M^2 \in [10, 15] {\rm{GeV}}^2$ with the selected threshold
parameter $s_0^{B_{0}}=35 {\rm{GeV}}^2$ as shown in Fig. \ref{form
factor of fplus}, where the ratio of twist-3 distribution amplitudes
in the total sum rules are also included for a comparison. Following
the same methods, we can also further evaluate all the form factors
$f_{+}(0)$, $f_{-}(0)$ and  $f_{T}(0)$ associating with the decay
modes $\bar{B}_0 \to a_0(1450) l \nu_l$, $\bar{B}_0 \to
K^{\ast}_0(1430) l \bar{l}$, $B_s \to K^{\ast}_0(1430) l \nu_l$, and
$B_s \to f_0(1500) l \bar{l}$, whose results have been collected in
Table \ref{di-fit B to a0(1450)}-\ref{di-fit Bs to f0(1500)}, where
we have combined the uncertainties from the variation of Borel
parameters, fluctuation of threshold value, errors of $b$ quark
mass, corrections from decay constants of the involved mesons as
well as uncertainties from the Gengenbauer moments in the
distribution amplitudes of scalar mesons. It can be observed that
the errors on the form factors are estimated within the level of 20
\% as expected by the general understanding of the theoretical
framework.

\begin{figure}[tb]
\begin{center}
\begin{tabular}{ccc}
\includegraphics[scale=0.6]{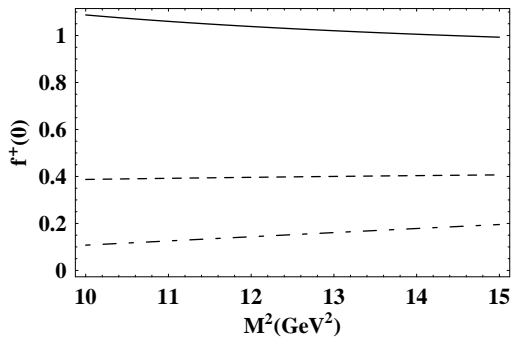}
\end{tabular}
\caption{The solid line denotes dependence of form factors $f_{+}$
at $q^2=0$ responsible for the decay of $\bar{B}_0 \to a_0(1450) l
\nu_l$ on the  Borel window $M_B^2 \in [10.0,15.0] {\mbox { GeV}^2}$
with the chosen threshold parameter $s_0^{B_{q_1}}=35 {\rm{GeV}}^2$.
The dashed and the dot-dashed lines are the ratio of contribution
from twist-3 distribution amplitudes and higher states of $B_{q_1}$
channel in the total sum rules respectively. }\label{form factor of
fplus}
\end{center}
\end{figure}

Next, we can further investigate the $q^2$ dependence of the form
factors $f_{+}(q^2)$, $f_{-}(q^2)$ and  $f_{T}(q^2)$  based on the
sum rules given in Eqs. (\ref{fplus}), (\ref{fminus}) and
(\ref{ftensor}).  One usually parameterize the form factors
$f_{i}(q^2) (i=+,-,T)$ in either the single pole form
\begin{eqnarray}
\label{form} f_{i}(q^2)={f_i(0) \over 1-a_i q^2/m_{B_{q_1}}^{2}},
\label{single-pole model of form factors}
\end{eqnarray}
or the double-pole form
\begin{eqnarray}
\label{form} f_{i}(q^2)={f_i(0) \over 1-a_i q^2/m_{B_{q_1}}^{2}+b_i
q^4/m_{B_{q_1}}^{4}}, \label{double-pole model of form factors}
\end{eqnarray}
in the whole kinematical region $0<q^2<(m_{B_{q_1}}-m_{S})^2$, while
non-perturbative parameters $a_i$ and $b_i$ can be fixed by the
magnitudes of form factors corresponding  to the small and
intermediate momentum transfer calculated in the LCSR approach. Our
results for the parameters $a_i$, $b_i$ accounting for the $q^2$
dependence of form factors $f_{+}$, $f_{-}$ and  $f_{T}$ are grouped
in Table \ref{di-fit B to a0(1450)}-\ref{di-fit Bs to f0(1500)},
where the values estimated in other works are also given for a
comparison.

\begin{table}[htb]
\caption{Numerical results for the parameters $f_i(0)$, $a_i$ and
$b_i$ involved in the (single) double-pole fit of form factors
(\ref{single-pole model of form factors}), (\ref{double-pole model
of form factors}) responsible for $\bar{B}_0 \to a_0(1450) l \nu_l$
decay up to the twist-3 distribution amplitudes of scalar mesons,
where the numbers derived in the covariant light-front quark model
\cite{Cheng:2003sm} are also collected for a comparison.}
\begin{tabular}{c c c c }
  \hline \hline
  & $\hspace{2 cm} f_i(0)$ & $ \hspace{2 cm} a_i$ & $\hspace{2 cm} b_i$
  \\ \hline
  $f_{+}$ & $\hspace{2 cm} 1.04^{+0.20}_{-0.20}$ & \hspace{2 cm} $0.98^{+0.08}_{-0.08}$ & \\
  & \hspace{2 cm}  0.52 \cite{Cheng:2003sm}& \hspace{2 cm}  1.57 \cite{Cheng:2003sm} & \hspace{2 cm}  0.70
  \cite{Cheng:2003sm} \\ \hline
  $f_{-}$ & $\hspace{2 cm} 0.077^{+0.014}_{-0.014}$ & \hspace{2 cm} $1.52^{+0.07}_{-0.12}$ &  \\
  \hline
  $f_{T}$ & $\hspace{2 cm} 0.66^{+0.13}_{-0.14}$ & \hspace{2 cm} $0.88^{+0.10}_{-0.09}$ &  \\
  \hline \hline
\end{tabular}
\label{di-fit B to a0(1450)}
\end{table}

\begin{table}[htb]
\caption{Numerical results for the parameters $f_i(0)$, $a_i$ and
$b_i$ involved in the (single) double-pole fit of form factors
(\ref{single-pole model of form factors}), (\ref{double-pole model
of form factors}) responsible for $\bar{B}_0 \to K^{\ast}_0(1430) l
\bar{l}$ decay up to the twist-3 distribution amplitudes of scalar
mesons, where the numbers derived in the covariant light-front quark
model \cite{Chen:2007na} and QCD sum rules approach
\cite{Aliev:2007rq} are also collected for a comparison.}
\begin{tabular}{c c c c }
  \hline \hline
  & $\hspace{2 cm} f_i(0)$ & $ \hspace{2 cm} a_i$ & $\hspace{2 cm} b_i$
  \\ \hline
  $f_{+}$ & $\hspace{2 cm} 0.97^{+0.20}_{-0.20}$ & \hspace{2 cm} $0.86^{+0.19}_{-0.18}$ & \\
  & \hspace{2 cm}  0.52 \cite{Chen:2007na}& \hspace{2 cm}  1.36 \cite{Chen:2007na} & \hspace{2 cm}
  0.86  \cite{Chen:2007na} \\
  & \hspace{2 cm} $0.62 \pm 0.16$ \cite{Aliev:2007rq} & \hspace{2 cm} 0.81 \cite{Aliev:2007rq}
  & \hspace{2 cm} -0.21 \cite{Aliev:2007rq} \\ \hline
  $f_{-}$ & $\hspace{2 cm} 0.073^{+0.02}_{-0.02}$ & \hspace{2 cm} $2.50^{+0.44}_{-0.47}$ & \hspace{2 cm} $1.82^{+0.69}_{-0.76}$ \\
  \hline
  $f_{T}$ & $\hspace{2 cm} 0.60^{+0.14}_{-0.13}$ & \hspace{2 cm} $0.69^{+0.26}_{-0.27}$ &  \\
  & \hspace{2 cm}  0.34 \cite{Chen:2007na}& \hspace{2 cm}  1.64 \cite{Chen:2007na} & \hspace{2 cm}
  1.72  \cite{Chen:2007na} \\
  & \hspace{2 cm} $0.26 \pm 0.07$ \cite{Aliev:2007rq} & \hspace{2 cm} 0.41 \cite{Aliev:2007rq}
  & \hspace{2 cm} -0.32 \cite{Aliev:2007rq} \\
  \hline \hline
\end{tabular}
\label{di-fit B to Kstar0(1430)}
\end{table}

\begin{table}[htb]
\caption{Numerical results for the parameters $f_i(0)$, $a_i$ and
$b_i$ involved in the (single) double-pole fit of form factors
(\ref{single-pole model of form factors}), (\ref{double-pole model
of form factors}) responsible for $B_s \to K^{\ast}_0(1430) l \nu_l$
decay up to the twist-3 distribution amplitudes of scalar mesons,
where the numbers derived in the  QCD sum rules approach
\cite{Yang:2005bv} are also collected for a comparison.}
\begin{tabular}{c c c c }
  \hline \hline
  & $\hspace{2 cm} f_i(0)$ & $ \hspace{2 cm} a_i$ & $\hspace{2 cm} b_i$
  \\ \hline
  $f_{+}$ & $\hspace{2 cm} 0.83^{+0.26}_{-0.13}$ & \hspace{2 cm} $0.93^{+0.20}_{-0.06}$ & \\
  & \hspace{2 cm}  $0.48 \pm 0.20$ \cite{Yang:2005bv} & \hspace{2 cm}  $1.25^{+0.07}_{-0.06}$ \cite{Yang:2005bv} &   \\ \hline
  $f_{-}$ & $\hspace{2 cm} 0.071^{+0.02}_{-0.02}$ & \hspace{2 cm} $2.46^{+0.36}_{-0.39}$ & \hspace{2 cm} $1.72^{+0.59}_{-0.64}$ \\
  \hline
  $f_{T}$ & $\hspace{2 cm} 0.52^{+0.18}_{-0.08}$ & \hspace{2 cm} $0.77^{+0.10}_{-0.07}$ &  \\
  \hline \hline
\end{tabular}
\label{di-fit Bs to Kstar0(1430)}
\end{table}

\begin{table}[htb]
\caption{Numerical results for the parameters $\xi_i(0)$, $a_i$ and
$b_i$ involved in the (single) double-pole fit of form factors
(\ref{single-pole model of form factors}), (\ref{double-pole model
of form factors}) responsible for $B_s \to f_0(1500) l \bar{l}$
decay up to the twist-3 distribution amplitudes of scalar mesons.}
\begin{tabular}{c c c c }
  \hline \hline
  & $\hspace{2 cm} f_i(0)$ & $ \hspace{2 cm} a_i$ & $\hspace{2 cm} b_i$
  \\ \hline
  $f_{+}$ & $\hspace{2 cm} 0.86^{+0.15}_{-0.15}$ & \hspace{2 cm} $1.17^{+0.06}_{-0.05}$ &  \\ \hline
  $f_{-}$ & $\hspace{2 cm} 0.056^{+0.015}_{-0.015}$ & \hspace{2 cm} $1.94^{+0.48}_{-0.85}$ & \hspace{2 cm} $0.52^{+0.89}_{-1.6}$ \\
  \hline
  $f_{T}$ & $\hspace{2 cm} 0.56^{+0.10}_{-0.11}$ & \hspace{2 cm} $1.09^{+0.08}_{-0.07}$ &  \\
  \hline \hline
\end{tabular}
\label{di-fit Bs to f0(1500)}
\end{table}

\newpage

\section{Decay Rate and Polarization Asymmetry}

With the transition form factors derived, one can proceed to perform
the calculations on some interesting observables in phenomenology,
such as decay rate, polarization asymmetry. In particular, the
forward-backward asymmetry for the decay modes $\bar{B}_0 \to
K^{\ast}_0(1430) l \bar{l}$ and $B_s \to f_0(1500) l \bar{l}$ is
exactly equal to zero in the SM \cite{Belanger,Geng} due to the
absence of scalar-type coupling between the lepton pair, which serve
as a valuable ground to test the SM precisely as well as bound its
extensions stringently.

The semi-leptonic decay $\bar{B}_0 \to K^{\ast}_0(1430) l \bar{l}$
is induced by flavor-changing neutral current. The differential
decay width of $\bar{B}_0 \to K^{\ast}_0(1430) l \bar{l}$ in the
rest frame of $\bar{B}_0$ meson can be written as \cite{PDG}
\begin{equation}
{d\Gamma(\bar{B}_0 \to K^{\ast}_0(1430) l \bar{l}) \over d q^2} ={1
\over (2 \pi)^3} {1 \over 32 m_{\bar{B}_0}} \int_{u_{min}}^{u_{max}}
|{\widetilde{M}}_{\bar{B}_0 \to K^{\ast}_0(1430) l \bar{l}}|^2 du,
\label{differential decay width}
\end{equation}
where $u=(p_{K^{\ast}_0(1430)}+p_{l})^2$ and
$q^2=(p_{l}+p_{\bar{l}})^2$; $p_{K^{\ast}_0(1430)}$, $p_{l}$ and
$p_{\bar{l}}$ are the four-momenta vectors of $K^{\ast}_0(1430)$,
$l$ and $\bar{l}$ respectively; $|{\widetilde{M}}_{\bar{B}_0 \to
K^{\ast}_0(1430) l \bar{l}}|^2$ is the squared decay amplitude after
integrating over the angle between the $l$ and $K^{\ast}_0(1430)$
baryon. The upper and lower limits of $u$ are given by
\begin{eqnarray}
u_{max}&=&(E^{\ast}_{K^{\ast}_0(1430)}+E^{\ast}_{l})^2-(\sqrt{E_{K^{\ast}_0(1430)}^{\ast
2}-m_{K^{\ast}_0(1430)}^2}-\sqrt{E_l^{\ast 2}-m_l^2})^2, \nonumber\\
u_{min}&=&(E^{\ast}_{K^{\ast}_0(1430)}+E^{\ast}_{l})^2-(\sqrt{E_{K^{\ast}_0(1430)}^{\ast
2}-m_{K^{\ast}_0(1430)}^2} +\sqrt{E_l^{\ast 2}-m_l^2})^2;
\end{eqnarray}
where $E^{\ast}_{K^{\ast}_0(1430)}$ and $E^{\ast}_{l}$ are the
energies of $K^{\ast}_0(1430)$ and $l$ in the rest frame of lepton
pair and can be determined as
\begin{equation}
E^{\ast}_{K^{\ast}_0(1430)}= {m_{\bar{B}_0}^2
-m_{K^{\ast}_0(1430)}^2 -q^2 \over 2 \sqrt{q^2}}, \hspace {1 cm}
E^{\ast}_{l}={q^2\over 2\sqrt{q^2}}.
\end{equation}
Collecting everything together, we can arrive at  the general
expression of differential decay rate for $ B_{q^{\prime
}}\rightarrow Sl\bar{l}$
 as \cite{Chen:2007na}:%
\begin{equation}
\frac{d\Gamma \left( B_{q^{\prime }}\rightarrow Sl\bar{l}\right) }{ds'}=\frac{%
G_{F}^{2}\left| V_{tb}V_{ts}\right| ^{2}m_{B}^{5}\alpha _{em}^{2}}{%
1536 \pi ^{5}}\left( 1-\frac{4r_{l}}{s'}\right) ^{1/2}\varphi _{S}^{1/2}%
\left[ \left( 1+\frac{2r_{l}}{s'}\right) \alpha _{S}+r_{l}\delta
_{S}\right] \label{drate}
\end{equation}%
where%
\begin{eqnarray*}
s' &=&q^{2}/m_{B}^{2}\text{, \ \ \ \ \ \ \
}r_{l}=m_{l}^{2}/m_{B}^{2}\text{,
\ \ \ \ \ \ \ }r_{S}=m_{S}^{2}/m_{B}^{2}\text{,} \\
\varphi _{S} &=&\left( 1-r_{S}\right) ^{2}-2s\left( 1+r_{S}\right) +s^{2}%
\text{,} \\
\alpha _{S} &=&\varphi _{S}\left( \left| C_{9}^{eff}{f_{+}\left(
q^{2}\right) \over 2} -2\frac{C_{7}f_{T}\left( q^{2}\right)
}{1+\sqrt{r_{S}}}\right| ^{2}+\left|
C_{10}{f_{+}\left( q^{2}\right) \over 2}\right| ^{2}\right) , \\
\delta _{S} &=&6\left| C_{10}\right| ^{2}\left\{ \left[ 2\left(
1+r_{S}\right) -s\right] \left| {f_{+}\left( q^{2}\right) \over 2}
\right| ^{2}+ \left( 1-r_{S}\right) {\rm{Re}} \left[ f_{+}\left(
q^{2}\right) (f_{-}(q^2)-{f_{+}\left( q^{2}\right) \over 2})^{\ast}
\right] +s\left| f_{-}(q^2)-{f_{+}\left( q^{2}\right) \over
2}\right| ^{2}\right\} .
\end{eqnarray*}%

The invariant dilepton mass distribution for $\bar{B}_0 \to
K^{\ast}_0(1430) l \bar{l}$  as the functions of squared momentum
transfer $q^2$ are presented in Fig. \ref{differential decay width}.
In the same way, we can also estimate the decay rates of $\bar{B}_0
\to K^{\ast}_0(1430) l \bar{l}$, $B_s \to K^{\ast}_0(1430) l
\bar{\nu}_l$ and $B_s \to f_0(1500) l \bar{l}$ with $l=e, \mu, \tau$
based on the form factors calculated in light-cone sum rules. The
results of the total decay width corresponding to these decay modes
are grouped in Table \ref{results of decay width}, where the results
obtained in other frameworks are also presented for a comparison. As
can be observed, the decay rates of the electron- and muon- pair
final states  are practically the same, while the decay rate of
tauon-pair  channel is much smaller due to the heavily suppressed
phase space.

\begin{figure}[tb]
\begin{center}
\begin{tabular}{ccc}
\includegraphics[scale=0.6]{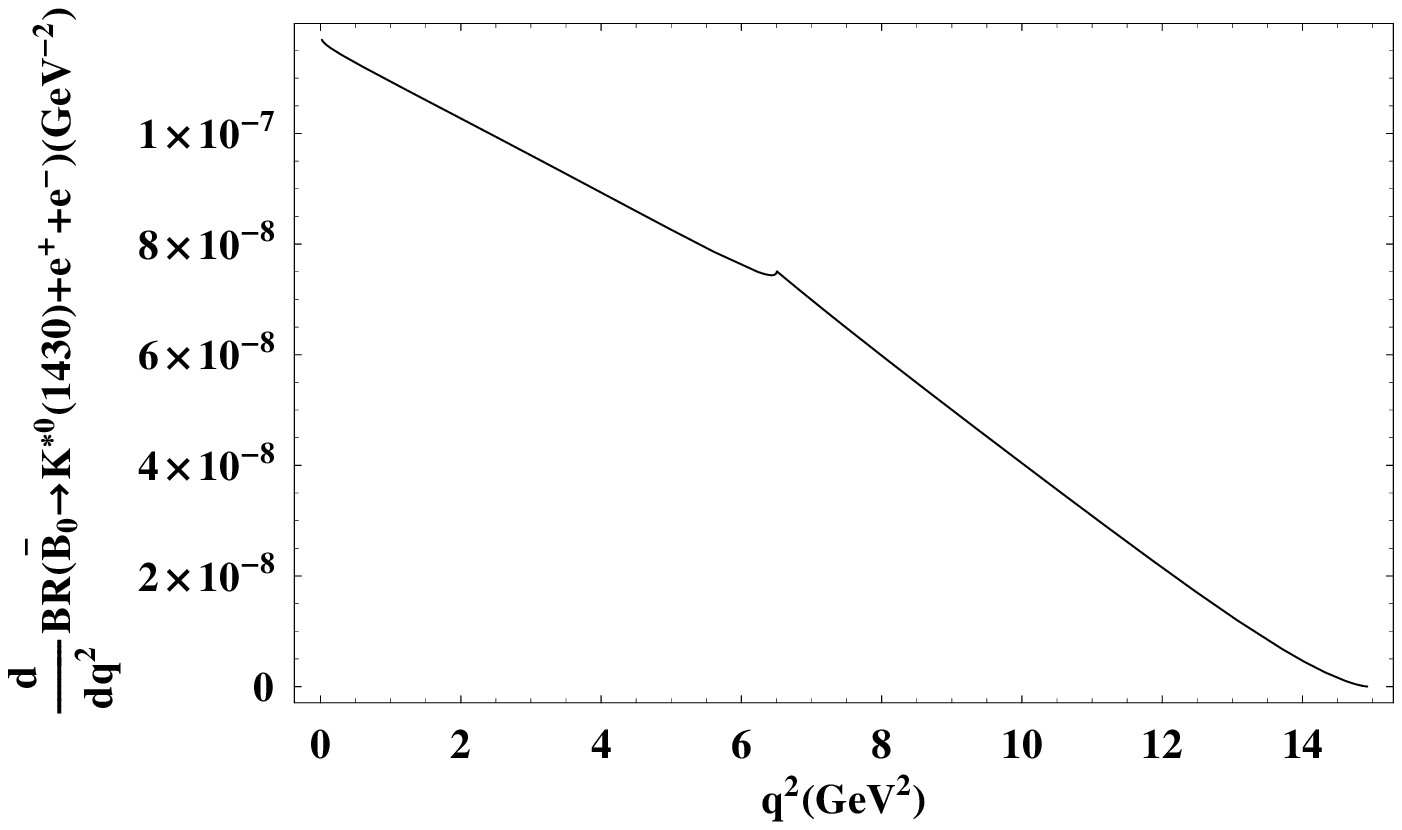}
\includegraphics[scale=0.6]{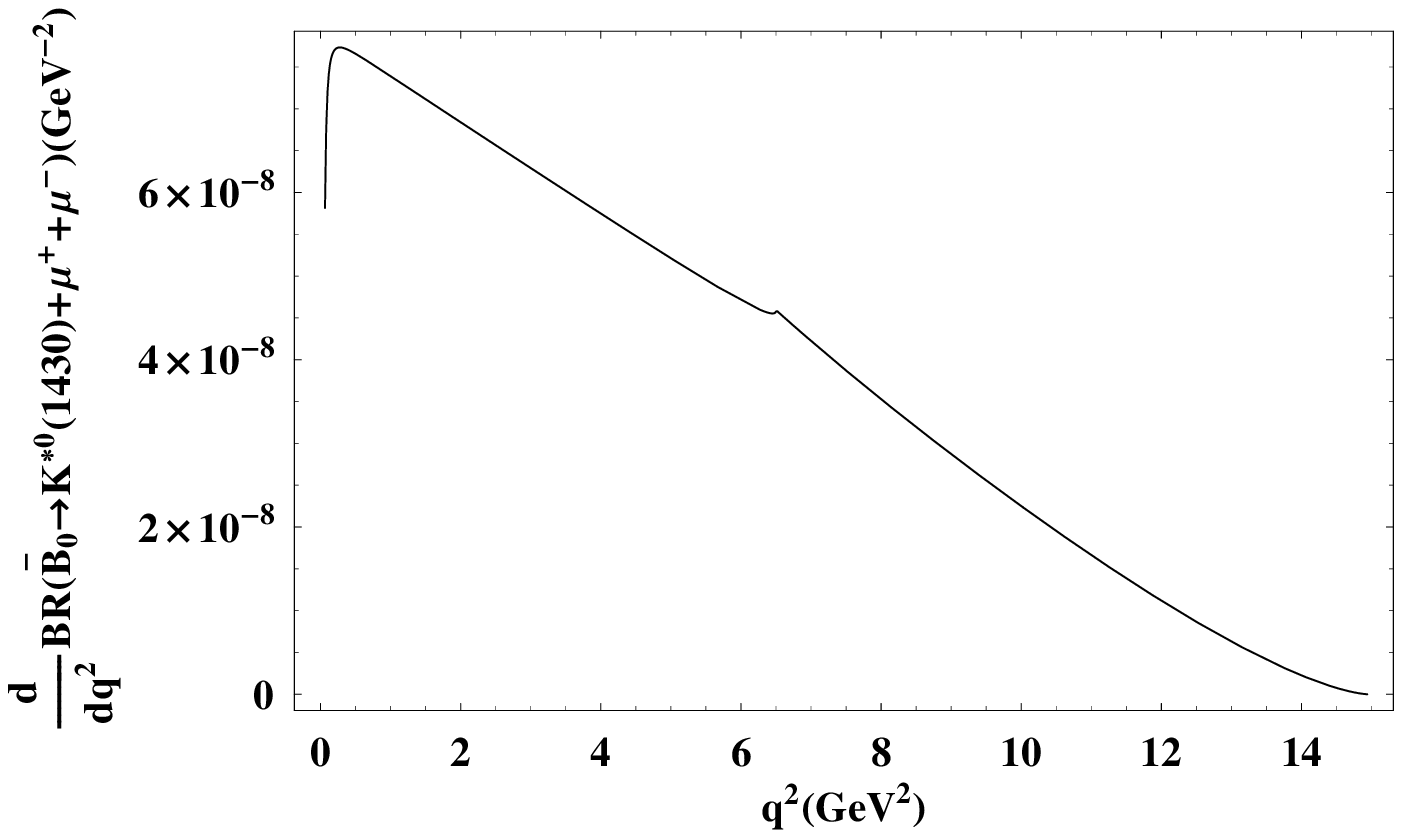}\\
\includegraphics[scale=0.6]{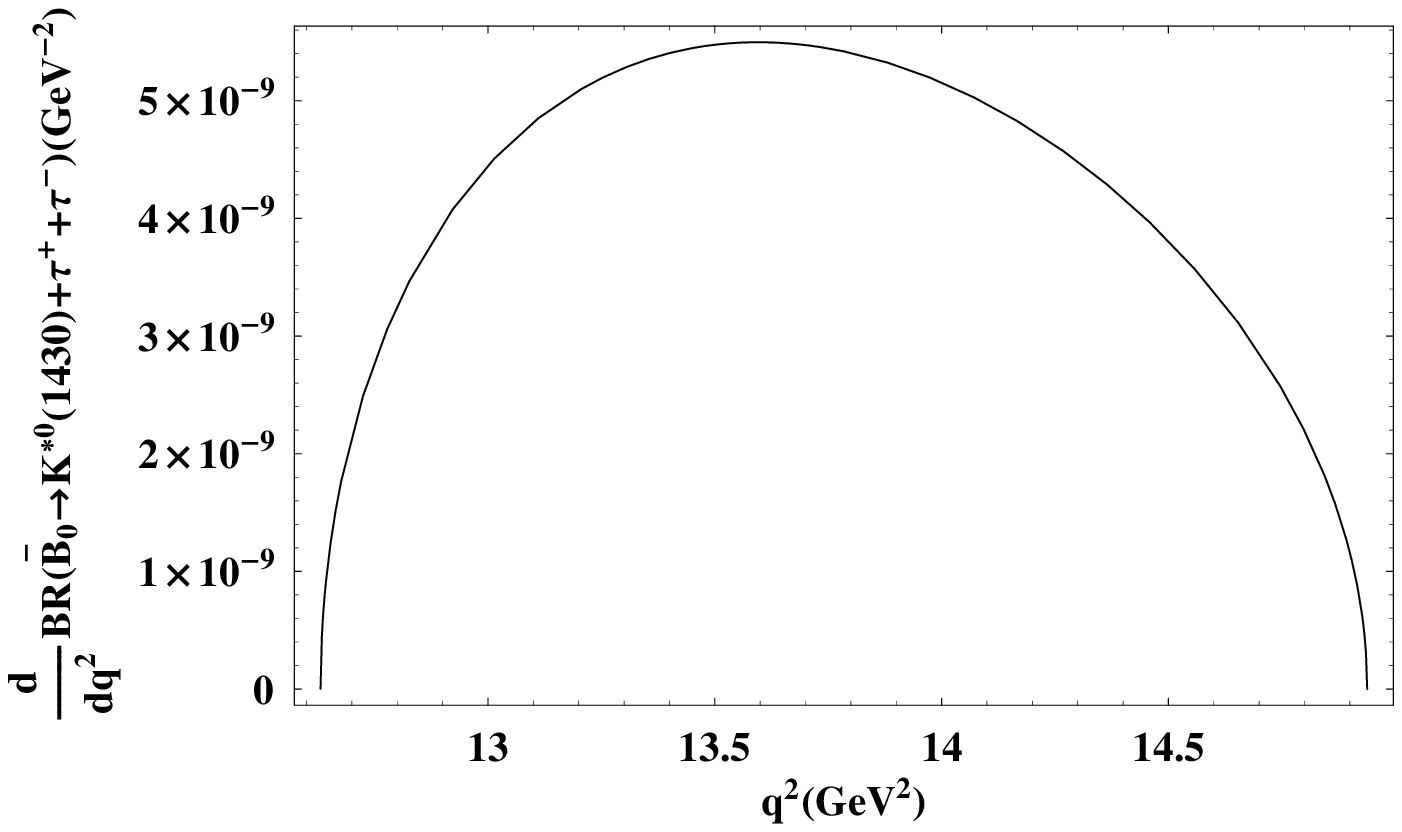}
\end{tabular}
\caption{The invariant dilepton mass distributions for $\bar{B}_0
\to K^{\ast}_0(1430) e^{+} e^{-}$, $\bar{B}_0 \to K^{\ast}_0(1430)
\mu^{+} \mu^{-}$ and $\bar{B}_0 \to K^{\ast}_0(1430) \tau^{+}
\tau^{-}$ as functions of squared momentum transfer $q^2$ based on
light-cone QCD sum rules. }\label{differential decay width}
\end{center}
\end{figure}

\begin{table}[htb]
\caption{Numerical results for the total decay width of $\bar{B}_0
\to a_0(1450) l \bar{\nu}_l$, $\bar{B}_0 \to K^{\ast}_0(1430) l
\bar{l}$, $B_s \to K^{\ast}_0(1430) l \bar{\nu}_l$ and $B_s \to
f_0(1500) l \bar{l}$ with $l=e, \mu, \tau$ in the light-cone sum
rules approach, together with  the numbers estimated in QCD sum
rules \cite{Aliev:2007rq,Yang:2005bv} and light-front quark model
\cite{Chen:2007na}.}
\begin{tabular}{c c c c c }
  \hline   \hline
   & \hspace{1 cm} $\bar{B}_0 \to a_0(1450) e \bar{\nu}_e$ & \hspace{1 cm} $\bar{B}_0 \to K^{\ast}_0(1430) e^{+} e^{-}$
   & \hspace{1 cm} $B_s \to K^{\ast}_0(1430)e \bar{\nu}_e $ & \hspace{1 cm} $B_s \to f_0(1500) e^{+} e^{-}$ \\
  LCSR & \hspace{1 cm} $1.8^{+0.9}_{-0.6} \times 10^{-4}$ & \hspace{1 cm} $5.7^{+3.4}_{-2.4} \times 10^{-7}$ & \hspace{1 cm} $1.3^{+1.3}_{-0.4} \times 10^{-4}$ & \hspace{1 cm} $5.3^{+2.3}_{-1.8} \times 10^{-7}$ \\
  LFQM& & \hspace{1 cm} $1.63 \times 10^{-7}$ \cite{Chen:2007na}& & \\
  QCDSR&  & \hspace{1 cm} $(2.09 \sim 2.68) \times 10^{-7}$\cite{Aliev:2007rq} & \hspace{1 cm} $3.6^{+3.8}_{-2.4} \times 10^{-5}$ \cite{Yang:2005bv}  &  \\
 \hline   \hline
  & \hspace{1 cm} $\bar{B}_0 \to a_0(1450) \mu \bar{\nu}_\mu$ & \hspace{1 cm} $\bar{B}_0 \to K^{\ast}_0(1430) \mu^{+} \mu^{-}$
   & \hspace{1 cm} $B_s \to K^{\ast}_0(1430)  \mu \bar{\nu}_\mu$ & \hspace{1 cm} $B_s \to f_0(1500) \mu^{+} \mu^{-}$ \\
 LCSR & \hspace{1 cm} $1.8^{+0.9}_{-0.7} \times 10^{-4}$ & \hspace{1 cm} $5.6^{+3.1}_{-2.3} \times 10^{-7}$ & \hspace{1 cm} $1.3^{+1.2}_{-0.4} \times 10^{-4}$ & \hspace{1 cm} $5.2^{+2.3}_{-1.7} \times 10^{-7}$\\
 LFQM  & & \hspace{1 cm} $1.62 \times 10^{-7}$ \cite{Chen:2007na}&  & \\
 QCDSR&  & \hspace{1 cm} $(2.07 \sim 2.66) \times 10^{-7}$\cite{Aliev:2007rq} & &  \\
  \hline   \hline
  & \hspace{1 cm} $\bar{B}_0 \to a_0(1450) \tau \bar{\nu}_\tau$ & \hspace{1 cm} $\bar{B}_0 \to K^{\ast}_0(1430) \tau^{+} \tau^{-}$
   & \hspace{1 cm} $B_s \to K^{\ast}_0(1430)\tau \bar{\nu}_\tau $ & \hspace{1 cm} $B_s \to f_0(1500)  \tau^{+} \tau^{-}$ \\
  LCSR & \hspace{1 cm} $6.3^{+3.4}_{-2.5} \times 10^{-5}$ & \hspace{1 cm} $9.8^{+12.4}_{-5.5} \times 10^{-9}$  & \hspace{1 cm} $5.2^{+5.7}_{-1.8} \times 10^{-5}$  & \hspace{1 cm} $1.2^{+0.8}_{-0.5} \times 10^{-8}$  \\
  LFQM& & \hspace{1 cm} $2.86 \times 10^{-9}$ \cite{Chen:2007na}&  & \\
  QCDSR&  & \hspace{1 cm} $(1.70  \sim 2.20) \times 10^{-9}$\cite{Aliev:2007rq} & &  \\
  \hline   \hline
\end{tabular}
\label{results of decay width}
\end{table}

\begin{figure}[tb]
\begin{center}
\begin{tabular}{ccc}
\includegraphics[scale=0.6]{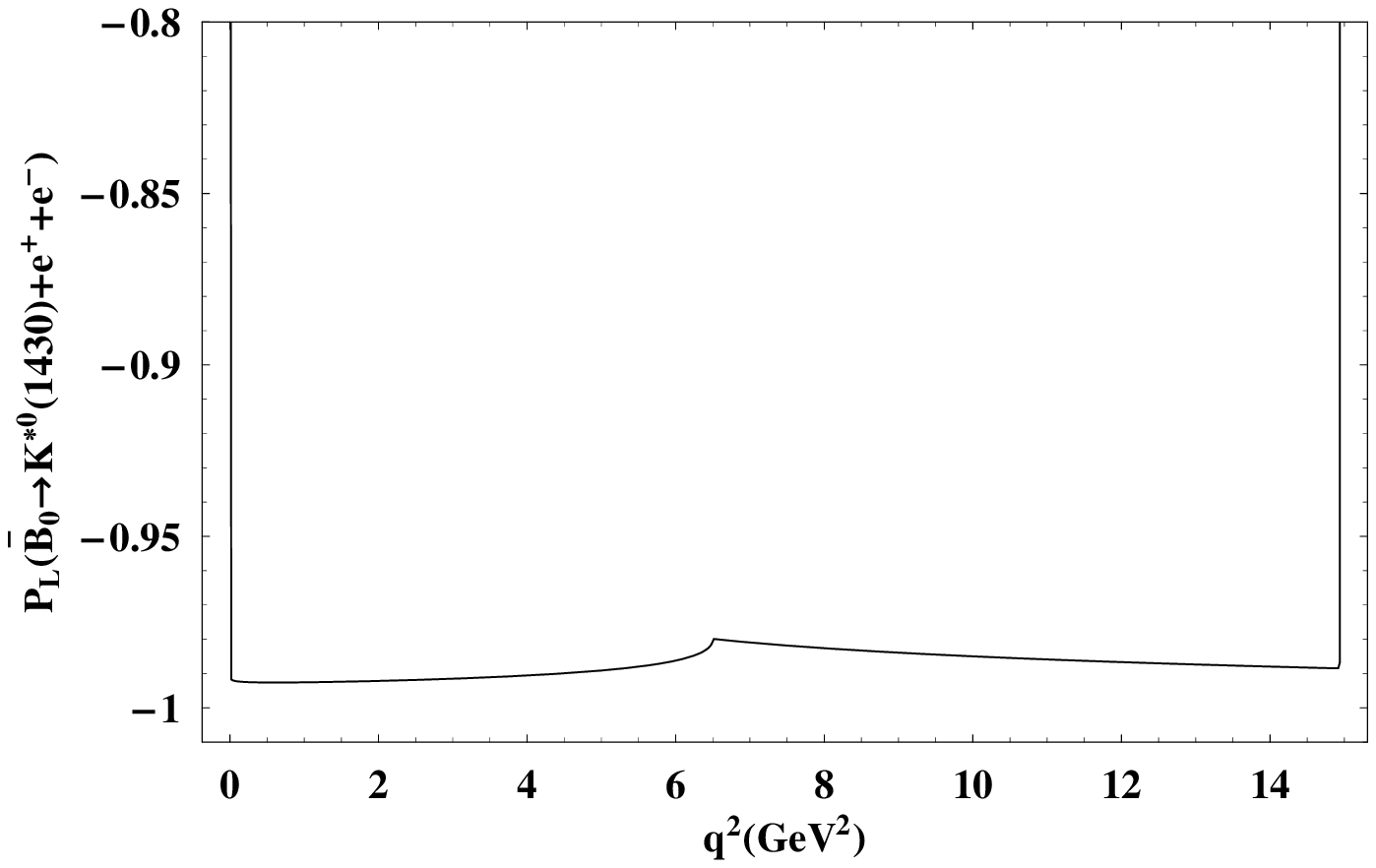}
\includegraphics[scale=0.6]{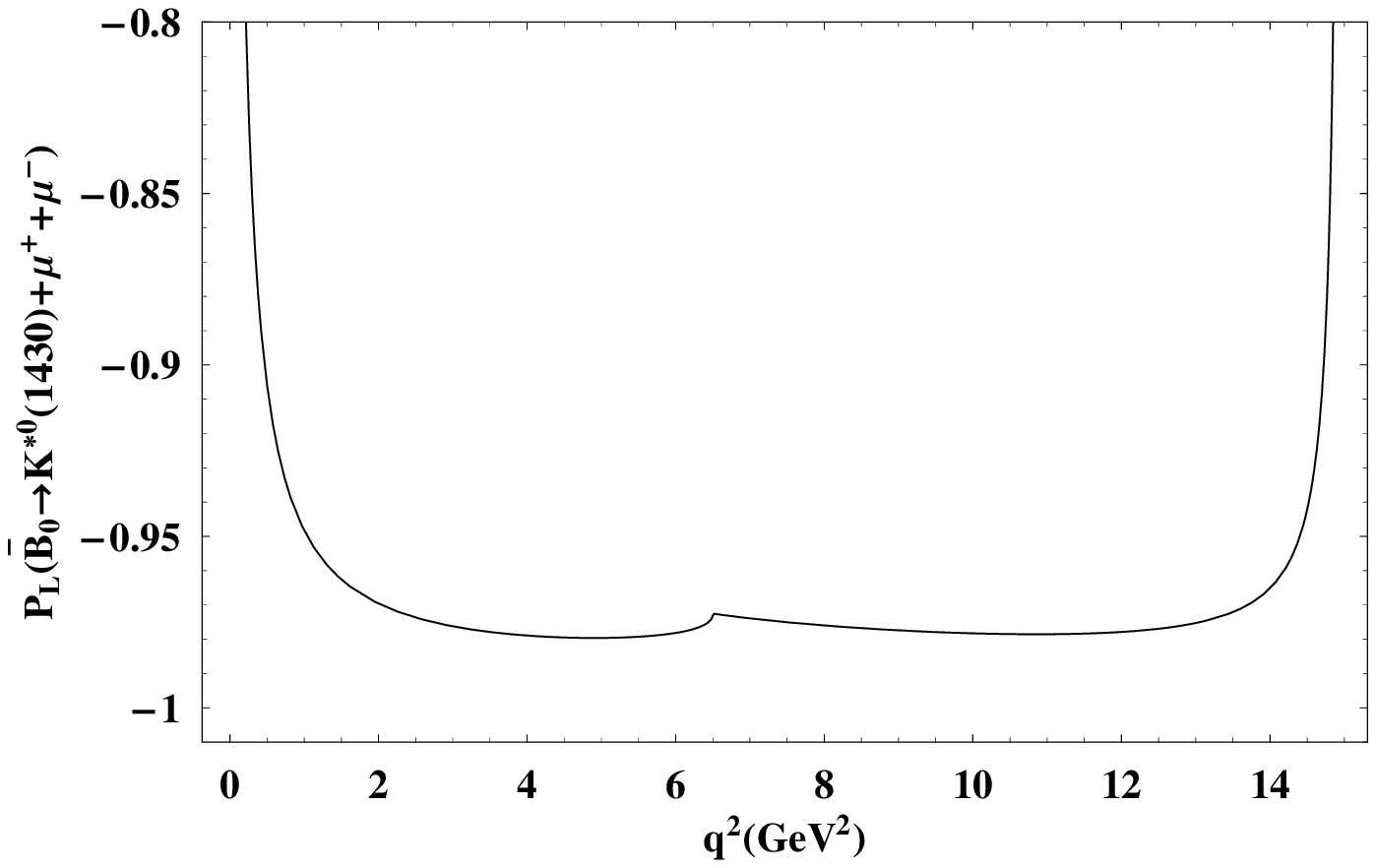}\\
\includegraphics[scale=0.6]{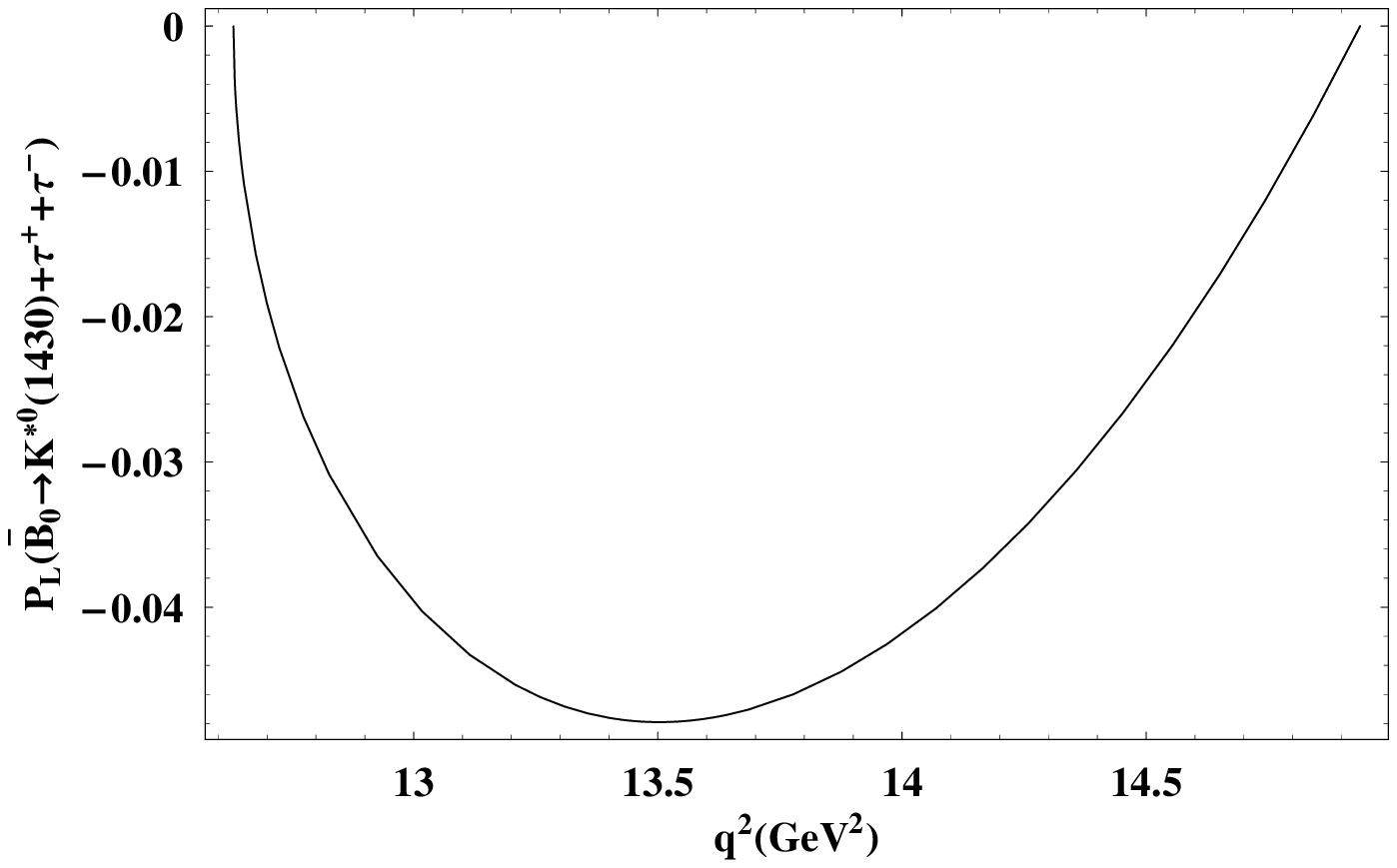}
\end{tabular}
\caption{Lepton polarization asymmetries for $\bar{B}_0 \to
K^{\ast}_0(1430) e^{+} e^{-}$, $\bar{B}_0 \to K^{\ast}_0(1430)
\mu^{+} \mu^{-}$ and $\bar{B}_0 \to K^{\ast}_0(1430) \tau^{+}
\tau^{-}$ as functions of squared momentum transfer $q^2$ based on
light-cone QCD sum rules. }\label{lepton polarization asymmetries}
\end{center}
\end{figure}

Another interesting observable in the decay of $\bar{B}_0 \to
K^{\ast}_0(1430) l \bar{l} $  is the polarization asymmetry of the
final state charged leptons, which is very helpful to extract the
information on the spin of them. The four-spin vector $s^{\mu}$ of a
lepton can be  defined in its rest frame as
\begin{eqnarray}
(s^{\mu})_{r.s.}=(0, \,\, \hat{\bf{\xi}}). \label{spin vector of
lambda baryon 1}
\end{eqnarray}
The unit vector along the longitudinal direction of the lepton
polarization is given by
\begin{eqnarray}
\hat{e}_L={\mathbf{p}_{l} \over |\mathbf{p}_{l}|}.
\end{eqnarray}
In this work, we mainly concentrate on the longitudinal  lepton
polarization asymmetry that can be defined as
\begin{equation}
P_{L}(s')=\frac{\frac{d\Gamma \left( \hat{e}_L \hat{\xi}=1\right)
}{ds'}-\frac{d\Gamma
\left( \hat{e}_L \hat{\xi}=-1 \right) }{ds'}}{\frac{d\Gamma \left( \hat{e}_L \hat{\xi}=1\right) }{ds'}+%
\frac{d\Gamma \left( \hat{e}_L \hat{\xi}=-1\right) }{ds'}},
\label{pasymmetry}
\end{equation}
which is a parity-odd but CP-even observable similar to the
forward-backward asymmetry. The manifest expression for the
longitudinal polarization asymmetry $P_L$ in $B_{q^{\prime
}}\rightarrow Sl\bar{l}$  is derived as \cite{Chen:2007na}
\begin{equation}
P_{L}(s')=\frac{2\left( 1-\frac{4r_{l}}{s'}\right) ^{1/2}}{\left( 1+\frac{2r_{l}}{%
s'}\right) \alpha _{S}+r_l\delta _{S}} {\rm{Re}} \left[ \varphi
_{S}\left(
C_{9}^{eff}{f_{+}\left( q^{2}\right) \over 2} -2\frac{C_{7}f_{T}\left( q^{2}\right) }{%
1+\sqrt{r_{S}}}\right) \left( C_{10}{f_{+}\left( q^{2}\right) \over 2} \right) ^{\ast }%
\right]   \label{pformula}
\end{equation}
It has been shown that this asymmetry is insensitive to the form
factors in the massless limit for the lepton and can be approximated
by
\begin{eqnarray}
P_L(s')={2 {\rm {Re}} [C_9^{eff} C_{10}^{\ast}]  \over |C_9^{eff}|^2
+|C_{10}|^2} + O(C_7) \simeq -1,
\end{eqnarray}
in view of the smallness of Wilson coefficient $C_7$ compared with
$C_9^{eff}$ and $C_{10}$. The distribution of the longitudinal
polarization asymmetry $P_L$ in $\bar{B}_0 \to K^{\ast}_0(1430) l
\bar{l} $ as a function of $q^2$ are presented in Fig. \ref{lepton
polarization asymmetries}, from which we indeed find that
$P_{L}(\bar{B}_0 \to K^{\ast}_0(1430) e^{+} e^{-})$ and
$P_{L}(\bar{B}_0 \to K^{\ast}_0(1430) \mu^{+} \mu^{-})$ are close to
$-1$ except the end points region.

It is also useful to introduce the integrated longitudinal  lepton
polarization asymmetry $\langle A_{PL} \rangle$ in order to
characterize the typical value of longitudinal  lepton polarization
asymmetry
\begin{eqnarray}
\langle A_{PL} \rangle= \int_{s'_{min}}^{s'_{max}}A_{PL}(s') ds' ,
\end{eqnarray}
with $s'_{min}={4 m_l^2 /m_{B}^2}$ and $s'_{max}={(m_{B}^2-m_{S}^2)
/m_{B}^2}$. The numerical results of integrated longitudinal  lepton
polarization asymmetry have been grouped in Table \ref{results of
polarization asymmetry}, together with results in light-front quark
model. From this table, we can observe that our results for $\langle
A_{PL} \rangle$ are in good agreement with that given by light-front
quark model, which also indicates that this asymmetry is not
sensitive to the decay form factors.

\begin{table}[htb]
\caption{Numerical results of the integrated longitudinal  lepton
polarization asymmetry  for $\bar{B}_0 \to K^{\ast}_0(1430) l
\bar{l}$ and $B_s \to f_0(1500) l \bar{l}$ with $l=e, \mu, \tau$ in
the light-cone sum rules approach, , where the numbers estimated in
light-front quark model \cite{Chen:2007na} are also collected here.
}
\begin{tabular}{ccc}
  \hline  \hline
 & \hspace{1 cm} $\bar{B}_0 \to K^{\ast}_0(1430) e^{+} e^{-}$ & \hspace{1 cm} $B_s \to f_0(1500) e^{+} e^{-}$ \\
  $\langle A_{PL} \rangle$& \hspace{1 cm} $-0.99 \pm 0.0$ & \hspace{1 cm} $-0.99 \pm 0.0$ \\
  & \hspace{1 cm}  $-0.97$ \cite{Chen:2007na}& \\
  \hline  \hline
  & \hspace{1 cm} $\bar{B}_0 \to K^{\ast}_0(1430) \mu^{+} \mu^{-}$ & \hspace{1 cm} $B_s \to f_0(1500) \mu^{+} \mu^{-}$ \\
  $\langle A_{PL} \rangle$& \hspace{1 cm} $-0.96 \pm 0.0$ & \hspace{1 cm}  $-0.96 \pm 0.0$\\
  & \hspace{1 cm}  $-0.95$ \cite{Chen:2007na} & \\
  \hline  \hline
  & \hspace{1 cm} $\bar{B}_0 \to K^{\ast}_0(1430) \tau^{+} \tau^{-}$ & \hspace{1 cm} $B_s \to f_0(1500) \tau^{+} \tau^{-}$ \\
  $\langle A_{PL} \rangle$& \hspace{1 cm} $-0.03^{+0.00}_{-0.01}$ & \hspace{1 cm} $-0.04 \pm 0.0$ \\
  & \hspace{1 cm}  $-0.03$ \cite{Chen:2007na}& \\
  \hline  \hline
\end{tabular}
\label{results of polarization asymmetry}
\end{table}

\section{Conclusions}

Within the framework of light-cone sum rules, we analyze the form
factors responsible for semi-leptonic decays of $\bar{B}_0 \to
a_0(1450) l \bar{\nu}_l$, $\bar{B}_0 \to K^{\ast}_0(1430) l
\bar{l}$, $B_s \to K^{\ast}_0(1430) l \bar{\nu}_l$ and $B_s \to
f_0(1500) l \bar{l}$ with $l=e, \mu, \tau$ up to the twist-3
distribution amplitudes for the leading Fock state. Owing to the
strong coupling of scalar mesons to the scalar current, the form
factors associating with $B \to S$ transition are approximately
twice as large as that for the ones in the $B \to P$ case. The form
factors $f_{+}(q^2)$, $f_{-}(q^2)$ and $f_{T}(q^2)$ calculated in
this work verify the relations derived in the large recoil and heavy
quark limit.

Utilizing these form factors, we investigated the branching
fractions of $\bar{B}_0 \to a_0(1450) l \bar{\nu}_l$, $\bar{B}_0 \to
K^{\ast}_0(1430) l \bar{l}$, $B_s \to K^{\ast}_0(1430) l
\bar{\nu}_l$ and $B_s \to f_0(1500) l \bar{l}$. The magnitudes of
$BR(\bar{B}_0 \to a_0(1450) l \bar{\nu}_l)$ and $BR(B_s \to
K^{\ast}_0(1430) l \bar{\nu}_l)$ can arrive of the order $10^{-4}$,
while the branching ratios of $\bar{B}_0 \to K^{\ast}_0(1430) l
\bar{l}$ and $B_s \to f_0(1500) l \bar{l}$ are of the order $10^{-8}
\sim 10^{-7}$, which can marginally observed in the future
experiments. The longitudinal lepton polarization asymmetries for
$\bar{B}_0 \to K^{\ast}_0(1430) l \bar{l}$ and $B_s \to f_0(1500) l
\bar{l}$ are also considered in the SM. Our results for the
asymmetry are in good agreement with that given by the light front
quark model. The averaged asymmetries $\langle A_{PL} \rangle$ for
the final states including $e^+ e^-$ and $\mu^+ \mu^-$ are almost
equal to $-1$ except the end points region. However, the tau lepton
polarization asymmetries are remarkably small and not measurable due
to the efficiency for the detectability of the tauon. The
theoretical predictions on the production properties presented in
this work are very helpful to clarify the inner structures of scalar
mesons as well as understanding the dynamics of strong interactions.

\section*{Acknowledgements}
This work is partly supported by National Science Foundation of
China under Grant No.10735080 and  10625525. The authors would like
to thank  T.M. Aliev and C.H. Chen for helpful discussions.

\end{document}